\documentclass[prb,twocolumn,showpacs]{revtex4}
\usepackage{graphicx,amsmath,amssymb}
\begin{document}

\renewcommand{\Re}{\mathop{\mathrm{Re}}}
\renewcommand{\Im}{\mathop{\mathrm{Im}}}
\renewcommand{\b}[1]{\mathbf{#1}}
\renewcommand{\c}[1]{\mathcal{#1}}
\renewcommand{\u}{\uparrow}
\renewcommand{\d}{\downarrow}
\renewcommand{\mod}{\mathop{\mathrm{mod}}}
\newcommand{\bsigma}{\boldsymbol{\sigma}}
\newcommand{\blambda}{\boldsymbol{\lambda}}
\newcommand{\tr}{\mathop{\mathrm{tr}}}
\newcommand{\sgn}{\mathop{\mathrm{sgn}}}
\newcommand{\sech}{\mathop{\mathrm{sech}}}
\newcommand{\diag}{\mathop{\mathrm{diag}}}
\newcommand{\half}{{\textstyle\frac{1}{2}}}
\newcommand{\sh}{{\textstyle{\frac{1}{2}}}}
\newcommand{\ish}{{\textstyle{\frac{i}{2}}}}
\newcommand{\thf}{{\textstyle{\frac{3}{2}}}}

\title{Kondo lattice on the edge of a two-dimensional topological insulator}

\author{Joseph Maciejko}

\affiliation{Princeton Center for Theoretical Science \& Department of Physics, Princeton University, Princeton, New Jersey 08544, USA}

\date\today

\begin{abstract}
We revisit the problem of a single quantum impurity on the edge of a two-dimensional time-reversal invariant topological insulator and show that the zero temperature phase diagram contains a large local moment region for antiferromagnetic Kondo coupling which was missed by previous poor man's scaling treatments. The combination of an exact solution at the so-called decoupling point and a renormalization group analysis \`{a} la Anderson-Yuval-Hamann allows us to access the regime of strong electron-electron interactions on the edge and strong Kondo coupling. We apply similar methods to the problem of a regular one-dimensional array of quantum impurities interacting with the edge liquid. When the edge electrons are at half-filling with respect to the impurity lattice, the system remains gapless unless the Luttinger parameter of the edge is less than $1/2$, in which case two-particle backscattering effects drive the system to a gapped phase with long-range Ising antiferromagnetic order. This is in marked contrast with the gapped disordered ground state of the ordinary half-filled one-dimensional Kondo lattice.
\end{abstract}


\maketitle

\section{Introduction}

Topological insulators\cite{hasan2010,qi2011} are recently discovered states of quantum matter which are topologically distinct from conventional insulators such as diamond or silicon. Topological insulators are characterized by a bulk energy gap, just like conventional insulators, but support gapless boundary modes that are unusually robust to external perturbations. The two-dimensional (2D) quantum spin Hall (QSH) insulator\cite{kane2005,bernevig2006b,maciejko2011} is the first time-reversal invariant topological insulator to be experimentally observed,\cite{konig2007,roth2009,brune2011} following its theoretical prediction in HgTe quantum wells.\cite{Bernevig2006} Its 1D boundary modes form a gapless helical liquid in which a Kramers' pair of states with opposite spin polarization counterpropagate at a given edge. The helical liquid is itself a new 1D gapless state of matter protected by time-reversal symmetry,\cite{Wu2006,Xu2006} distinct from the conventional spinless and spinful Luttinger liquids.\cite{Giamarchi2003} As long as the 2D bulk gap does not close, the helical liquid is robust against potential scatterers of arbitrary strength, concentration, or degree of randomness, provided that the strength of repulsive electron-electron interactions in the helical liquid does not exceed a critical value which is finite,\cite{Wu2006} rather than infinitesimal as is the case for the ordinary spinless and spinful Luttinger liquids.\cite{Kane1992}

However, the helical liquid is not necessarily protected against magnetic impurities. Classical magnetic impurities, i.e., static magnetic moments, act on the helical liquid just as potential scatterers do on a spinless Luttinger liquid. In the presence of such impurities, infinitesimally weak electron-electron repulsive interactions are sufficient to renormalize the conductance of the helical liquid to zero at zero temperature. The physics is more subtle in the case of quantum impurities. Time-reversal symmetry allows for two types of such perturbations:\cite{Wu2006} dynamical, local magnetic moments coupled by magnetic exchange to the spin of neighboring edge electrons, and localized interaction centers which backscatter two edge electrons at a time. Those kinds of quantum impurities might occur in the HgTe QSH state due to potential inhomogeneities which can trap bulk electrons in a small region and force them to interact with the edge electrons. As has been argued previously,\cite{maciejko2009} such localized perturbations might account for the deviation of the observed longitudinal conductance from its predicted quantized value of $2e^2/h$ as well as its unusual temperature dependence.\cite{MarkusThesis} Experimental efforts are underway\cite{KoenigAPS2011} which might help test those predictions.

In addition to being a question of experimental relevance, the study of quantum impurities interacting with the helical edge modes of the QSH state acquires a broader theoretical significance in the context of the study of strong correlation effects in topological insulators, which is a topic of tremendous current interest.\cite{young2008,rachel2010,goryo2010,hohenadler2011,goryo2011,yamaji2011,wu2011,yu2011,griset2012,lee2011,zheng2011,hohenadler2012,hohenadler2011b,yoshida2011,jie2012} Recent quantum Monte Carlo studies\cite{zheng2011,hohenadler2012} of the Kane-Mele-Hubbard model indicate that as the strength $U$ of the on-site Hubbard interaction increases from zero to some critical value $U_c$, the 2D bulk remains paramagnetic and time-reversal invariant while the effective Luttinger parameter $K$ of the edge decreases from the noninteracting value $K=1$ to values $K<1/2$. This means that it is possible to reach a regime, at least numerically, where the helical edge liquid is strongly interacting, i.e., where $1-K\sim\mathcal{O}(1)$. However, previous studies\cite{Wu2006,maciejko2009} of a single quantum impurity interacting with the helical edge liquid are in fact perturbative in $1-K$, as will be seen. Those studies are also perturbative in the Kondo coupling $J$ between the impurity and the edge liquid. In order to further our understanding of strong correlation effects in topological insulators, it is desirable to revisit those studies and extend them beyond the weak coupling regime $1-K\ll 1$ and $\rho J\ll 1$ with $\rho$ the density of states of the helical liquid. Finally, it is natural to ask what happens when several quantum impurities are present along the edge. From an experimental point of view, impurities are not necessarily isolated and interimpurity coherence effects might play an important role in transport properties at low temperatures. From a theoretical point of view, one expects that effective interactions of the Ruderman-Kittel-Kasuya-Yosida (RKKY) type will be mediated between quantum impurities by the helical edge electrons,\cite{gao2009} and it is natural to ask what particular quantum phases will be formed by a collection of quantum impurities under the influence of such interactions.

In this paper, we revisit the problem of a single quantum impurity interacting with the edge of a QSH insulator (Sec.~\ref{sec:SingleImpurity}) and construct an improved zero temperature phase diagram in the $(K,J_z)$ plane where $J_z$ is the Kondo coupling for spins in the $z$ direction. We consider the original model of the QSH state as two copies of the quantum Hall state with opposite spin and chirality,\cite{kane2005,bernevig2006b} where the $z$ component of the total spin is conserved. Our revised phase diagram [Fig.~\ref{fig:fig2}(b)] differs markedly from that which is inferred from previous results [Fig.~\ref{fig:fig2}(a)], especially due to a newly found large portion of the phase diagram for antiferromagnetic $J_z$ which does not exhibit Kondo screening. We then generalize this problem to a regular 1D array of quantum impurities (Sec.~\ref{sec:KondoLattice}) for which we derive a zero temperature phase diagram (Fig.~\ref{fig:PhaseDiagramKL}). This phase diagram can be contrasted to that of the ordinary 1D Kondo lattice.\cite{tsunetsugu1997} The most striking difference is found at half-filling, where our system remains gapless for $K>1/2$ but becomes gapped and develops long-range Ising antiferromagnetic order for $K<1/2$, while the ordinary 1D Kondo lattice is gapped but has no long-range magnetic order.

\section{Revisiting the single-impurity problem}\label{sec:SingleImpurity}

Our first goal is to obtain a phase diagram for the single-impurity problem as a function of the Luttinger parameter $K$ of the helical liquid and the Kondo coupling $J_z$. As mentioned before, previous analyses\cite{Wu2006,maciejko2009} were perturbative in $1-K$ and in $J_z$, such that only the regions $1-K\ll 1$ and $\rho J_z\ll 1$ were accessible. In Sec.~\ref{sec:SingleImpDecoupling}, we describe the special ``decoupling'' line $\rho J_z=2K$ along which the Kondo Hamiltonian becomes exactly solvable. In Sec.~\ref{sec:SingleImpAYH}, we obtain the phase diagram of the single-impurity problem for all $J_z$ and $0<K<1$ using renormalization group (RG) equations which are perturbative in $J_\perp$ but exact in $1-K$ and in $J_z$. We derive those equations using a method which is equivalent to the Anderson-Yuval-Hamann procedure\cite{anderson1970} but simpler in its application.

The Hamiltonian of the single-impurity problem in the bosonized representation is\cite{Wu2006,maciejko2009}
\begin{align}\label{HSingleImp}
&H=H_\mathrm{TL}+H_z+H_\perp\nonumber\\
&\hspace{3mm}=\frac{v_F}{2}\int dx\left[K\Pi^2+\frac{1}{K}(\partial_x\phi)^2\right]-\frac{J_z a}{\sqrt{\pi}}S^z\Pi(0)\nonumber\\
&\hspace{10mm}+\frac{J_\perp a}{2\pi\xi}\left(S^+e^{i2\sqrt{\pi}\phi(0)}+\mathrm{H.c.}\right),
\end{align}
where $v_F$ is the Fermi velocity of the edge electrons, $K$ is their Luttinger parameter, $a$ is the size of the impurity, $\xi$ is the penetration length of the edge states into the bulk and acts as a short-distance cutoff, and $S^\pm=S^x\pm iS^y$ and $S^z$ are the spin-$\half$ operators for the impurity spin localized at $x=0$. The bosonic fields $\phi(x)$ and $\Pi(x)$ describe the low-energy degrees of freedom of the helical liquid and satisfy the equal-time canonical commutation relations $[\phi(x),\Pi(x')]=i\delta(x-x')$. The first term in Eq.~(\ref{HSingleImp}) is the Tomonaga-Luttinger Hamiltonian which describes the translationally invariant, unperturbed helical liquid in the absence of impurities. The second and third terms represent the anisotropic Kondo interaction. The helical liquid has no $SU(2)$ spin rotation symmetry, hence the Kondo interaction is generally anisotropic. The Hamiltonian (\ref{HSingleImp}) has two conserved charges, $Q_c=\int dx\frac{1}{\sqrt{\pi}}\partial_x\phi$ corresponding to the $U(1)_c$ electromagnetic gauge invariance and $Q_s=\half\int dx(-\frac{1}{\sqrt{\pi}})\Pi+S^z$ corresponding to the $U(1)_s$ or XY spin rotation symmetry. Although the $U(1)_s$ symmetry is not required by the topology of the QSH state,\cite{schmidt2011} it is present in the simplest model of the QSH state\cite{kane2005,bernevig2006b} as two copies of the quantum Hall state with opposite spin and chirality.

\subsection{Decoupling limit}\label{sec:SingleImpDecoupling}

We first consider a particular line in the space of coupling constants for which the Hamiltonian (\ref{HSingleImp}) becomes exactly solvable. A solution of this type was considered recently,\cite{tanaka2011} which corresponds to the well-known Toulouse limit\cite{toulouse1969} of the Kondo problem in which the Kondo Hamiltonian reduces to a noninteracting resonant level problem. The solution we are considering here is the ``decoupling limit'',\cite{Gogolin} in which the Kondo Hamiltonian reduces to a problem where the impurity effectively decouples from the conduction electrons. The decoupling limit corresponds in fact to the unitarity limit $\delta=\frac{\pi}{2}$ where $\delta$ is the scattering phase shift of the conduction electrons (see Appendix A). For a Kondo impurity embedded in a 3D metallic host, the scattering phase shift $\delta$ is given by $\tan\delta\propto\rho J_z$,\cite{anderson1969} hence the decoupling limit is unphysical because $\delta=\frac{\pi}{2}$ corresponds to an infinite coupling $J_z=\infty$. However, for a 1D metallic host the scattering phase shift is given by $\delta\propto\rho J_z$, hence the decoupling limit corresponds to a finite value of $J_z$ and is therefore physical.

We begin by introducing a unitary transformation\cite{Emery1992,maciejko2009} which we will use repeatedly in this paper. We define the unitary operator $U=e^{i\lambda\phi(0)S^z}$ under which the various fields transform as
\begin{align}
Uf(\phi(x),\partial_x\phi(x),\ldots)U^\dag&=f(\phi(x),\partial_x\phi(x),\ldots),\label{SingleImpPhiTransform}\\
U\Pi(x)^2U^\dag&=\Pi(x)^2-2\lambda S^z\Pi(0)\delta(x),\nonumber\\
US^\pm U^\dag&=S^\pm e^{i\lambda\phi(0)},\label{SingleImpSpmTransform}\\
US^zU^\dag&=S^z,\nonumber
\end{align}
where $f$ is any function of $\phi$ and its spatial derivatives. Using these relations we find that the Hamiltonian (\ref{HSingleImp}) transforms as
\begin{align}
\tilde{H}\equiv UHU^\dag&=H_\mathrm{TL}[\phi,\Pi]-\frac{\tilde{J}_z a}{\sqrt{\pi}}S^z\Pi(0)\nonumber\\
&\hspace{5mm}+\frac{J_\perp a}{2\pi\xi}\left(S^+e^{i(2\sqrt{\pi}+\lambda)\phi(0)}+\mathrm{H.c.}\right),\nonumber
\end{align}
where $\tilde{J}_za=J_za+\lambda\sqrt{\pi}v_FK$. We observe that the exponent of the vertex operator can be canceled by choosing $\lambda=-2\sqrt{\pi}$. Choosing $\lambda$ as such, we obtain $\tilde{J}_za=J_za-2\pi v_FK$, such that for the special value
\begin{align}\label{SingleImpDecouplingLimit}
\rho J_z=2K,
\end{align}
where $\rho=a/\pi v_F$ is the density of states of the helical liquid, the transformed Hamiltonian is simply
\begin{align}\label{SingleImpHDecoupled}
\tilde{H}=H_\mathrm{TL}[\phi,\Pi]+\frac{J_\perp a}{\pi\xi}\b{S}\cdot\hat{\b{e}},
\end{align}
where $\hat{\b{e}}=\hat{\b{x}}$. The impurity spin completely decouples from the conduction electrons, since the two terms in Eq.~(\ref{SingleImpHDecoupled}) commute. Since $[\b{S}\cdot\hat{\b{e}},\tilde{H}]=0$, the projection of $\b{S}$ onto $\hat{\b{e}}$ is a good quantum number under $\tilde{H}$. In the ground state of $\tilde{H}$, we therefore obtain $\b{S}\cdot\hat{\b{e}}=-\half\sgn J_\perp$. However, $\b{S}\cdot\hat{\b{e}}$ is \emph{not} a good quantum number under the original Hamiltonian $H$, because of the transformation law (\ref{SingleImpSpmTransform}). We have
\begin{align}
-\half\sgn J_\perp=\langle\b{S}\cdot\hat{\b{e}}\rangle_{\tilde{H}}=\langle U^\dag\b{S}U\cdot\hat{\b{e}}\rangle_H
=\langle\b{S}\cdot\hat{\b{e}}(\phi(0))\rangle_H,\nonumber
\end{align}
where we define the unit vector
\begin{align}\label{SingleImpUnitVectorPhi}
\hat{\b{e}}(\phi(0))\equiv\hat{\b{x}}\cos 2\sqrt{\pi}\phi(0)-\hat{\b{y}}\sin 2\sqrt{\pi}\phi(0).
\end{align}
The impurity spin is entirely in the $xy$ plane, and the angle that it makes with the $x$ axis is locked to the local charge density wave (CDW) phase $2\sqrt{\pi}\phi(0)$ of the helical edge electrons. The dynamics of the impurity is therefore entirely controlled by the Hamiltonian $H_\mathrm{TL}$ of the conduction electrons. The impurity correlation functions are easily obtained by making use of the unitary transformation $U$. The imaginary time transverse spin-spin correlation function at zero temperature is given at long times by
\begin{align}\label{SingleImpChiTransverseTime}
\chi_\perp(\tau)&=\langle T_\tau S^+(\tau)S^-(0)\rangle_H\nonumber\\
&=\langle T_\tau S^+(\tau)S^-(0)\rangle_{\tilde{H}}\left\langle e^{i\lambda[\phi(\tau)-\phi(0)]}\right\rangle_{\tilde{H}}\nonumber\\
&=\frac{1+e^{-\omega_\perp|\tau|}}{4(\Lambda|\tau|)^{2K}},
\end{align}
where $\Lambda=v_F/\xi$ is a high-energy cutoff, we define $\omega_\perp\equiv|J_\perp|a/\pi\xi$, and we have used the expression\cite{Giamarchi2003}
\begin{align}\label{TomonagaLuttingerPropagator}
\langle T_\tau\phi(x,\tau)\phi(0,0)\rangle_\mathrm{TL}=-\frac{K}{4\pi}\ln\left(\frac{x^2+(v_F\tau)^2+\xi^2}{\xi^2}\right),
\end{align}
for the propagator associated to the Tomonaga-Luttinger Hamiltonian, as well as the fact that
\begin{align}
\left\langle e^{i\lambda[\phi(\tau)-\phi(0)]}\right\rangle_{\tilde{H}}=\exp\left(\lambda^2\langle T_\tau\phi(\tau)\phi(0)\rangle_{\tilde{H}}\right),\nonumber
\end{align}
since $H_\mathrm{TL}$ is a quadratic boson Hamiltonian. The frequency-dependent transverse susceptibility $\chi''_\perp(\omega)$ is defined as the imaginary part of the Fourier transform of the retarded correlation function $\chi^R_\perp(t)=i\theta(t)\langle[S^+(t),S^-(0)]\rangle_H$, which can be obtained directly from Eq.~(\ref{SingleImpChiTransverseTime}) by analytic continuation in the time domain,\cite{TsvelikBook}
\begin{align}
\chi''_\perp(\omega)=\half\int_{-\infty}^\infty dt\,e^{i\omega t}\left[\chi_\perp^+(it)-\chi_\perp^-(it)\right],\nonumber
\end{align}
where $\chi_\perp^\pm$ are defined by $\chi_\perp(\tau)=\theta(\tau)\chi_\perp^+(\tau)+\theta(-\tau)\chi_\perp^-(\tau)$. We obtain at low frequencies
\begin{align}\label{SingleImpChiTransverse}
&\chi''_\perp(\omega)=\frac{\pi\Lambda^{-2K}}{4\Gamma(2K)}\nonumber\\
&\hspace{8mm}\times\left[|\omega|^{2K-1}+(|\omega|-\omega_\perp)^{2K-1}\theta(|\omega|-\omega_\perp)\right]\sgn\omega,
\end{align}
which is odd in frequency, as required for a bosonic spectral function. The longitudinal spin-spin correlation function is given by
\begin{align}\label{SingleImpChizzTime}
\chi_{z}(\tau)=\langle T_\tau S^z(\tau)S^z(0)\rangle_H&=\langle T_\tau S^z(\tau)S^z(0)\rangle_{\tilde{H}}\nonumber\\
&={\textstyle\frac{1}{4}}e^{-\omega_\perp|\tau|},
\end{align}
hence the frequency-dependent longitudinal susceptibility is given by
\begin{align}\label{SingleImpChizz}
\chi''_z(\omega)=\frac{\pi}{4}\delta(|\omega|-\omega_\perp)\sgn\omega,
\end{align}
and is entirely unaffected by the interactions in the helical liquid. Note that the results (\ref{SingleImpChiTransverse}) and (\ref{SingleImpChizz}) are not only exact in $J_z$ and $K$ on the decoupling line (\ref{SingleImpDecouplingLimit}), but they are also exact in $J_\perp$.

\subsection{Away from the decoupling limit: Anderson-Yuval-Hamann approach}\label{sec:SingleImpAYH}

The results Eq.~(\ref{SingleImpChiTransverse}) and (\ref{SingleImpChizz}) that we found for the impurity spin susceptibilities hold only in the decoupling limit (\ref{SingleImpDecouplingLimit}), which is a line in the plane of $K$ and $J_z$. In this section we derive renormalization group (RG) equations [Eq.~(\ref{SingleImpRGJz}) and (\ref{SingleImpRGJperp})] which will allow us to explore the phase diagram of the single-impurity problem away from that special line. RG equations for the single-impurity problem have been derived previously\cite{Wu2006} and read
\begin{align}
\frac{dJ_\perp}{d\ell}&=(1-K)J_\perp+\rho J_zJ_\perp,\label{SingleImpRGPerturbative1}\\
\frac{dJ_z}{d\ell}&=\rho J_\perp^2.\label{SingleImpRGPerturbative2}
\end{align}
Those equations were obtained using Anderson's poor man's scaling approach,\cite{anderson1970b} and as such are perturbative in both $J_\perp$ and $J_z$. The poor man's scaling approach considers the first term of Eq.~(\ref{HSingleImp}) as the unperturbed Hamiltonian, and the $J_z$ and $J_\perp$ terms as perturbations. This is reasonable because the first term of Eq.~(\ref{HSingleImp}) is the free boson Hamiltonian which is exactly solvable. However, the original approach of Anderson, Yuval, and Hamann\cite{anderson1970} in which the Kondo problem is viewed as a succession of X-ray edge problems is exact in the phase shift associated to $J_z$. In this approach, one considers the first two terms of Eq.~(\ref{HSingleImp}) as the unperturbed Hamiltonian. In the basis of eigenstates of $S^z$, this Hamiltonian is simply that of a free boson scattering off a potential impurity and is also exactly solvable in terms of scattering phase shifts. Therefore, it is not necessary to treat $J_z$ as a perturbation. As we will see, Eq.~(\ref{SingleImpRGPerturbative1}) and (\ref{SingleImpRGPerturbative2}) are also perturbative in $1-K$. This is because they were obtained using the poor man's scaling approach in the fermion language, where the interactions between edge electrons are treated perturbatively. We will use the bosonized description of the edge electrons, in which electron-electron interactions represented by the Luttinger parameter $K$ can be treated exactly. Our treatment is conceptually equivalent to the Anderson-Yuval-Hamann approach, but is made technically simpler by the use of bosonization techniques.\cite{Giamarchi2003}

The imaginary time action at zero temperature corresponding to the Hamiltonian~(\ref{HSingleImp}) is $S=S_0+S_\perp$ with $S_0=S_\mathrm{TL}+S_z+S_\mathrm{WZ}$ where
\begin{align}
S_\mathrm{TL}&=\frac{1}{K}\int\frac{d\omega}{2\pi}|\omega||\phi(\omega)|^2,\label{SingleImpSTL}\\
S_z&=-\frac{J_za}{\sqrt{\pi}v_F}\int_0^\infty d\tau\,S^zi\partial_\tau\phi,\label{SingleImpSz}\\
S_\perp&=\frac{J_\perp a}{2\pi\xi}\int_0^\infty d\tau\left(S^+e^{i2\sqrt{\pi}\phi}+\mathrm{c.c.}\right),\label{SingleImpSperp}
\end{align}
and $S_\mathrm{WZ}$ is the Wess-Zumino or Berry phase term for the impurity spin\cite{Fradkin} whose exact expression is not needed because we will revert to the operator formalism for the computation of impurity spin correlators. As our notation suggests, $S_0$ including the $J_z$ Kondo term is used as the unperturbed action. We have defined $\phi(\tau)\equiv\phi(x=0,\tau)$, and $S_\mathrm{TL}$ in Eq.~(\ref{SingleImpSTL}) is obtained from the full $(1+1)$-dimensional Tomonaga-Luttinger action by integrating out $\phi(x\neq 0,\tau)$.\cite{Kane1992}

We use the standard Wilsonian RG procedure in which we define slow fields $\phi_<$, $\b{S}_<$ and fast fields $\phi_>$, $\b{S}_>$,
\begin{align}
\phi_<(\tau)&=\int_{-\Lambda/b}^{\Lambda/b}\frac{d\omega}{2\pi}e^{-i\omega\tau}\phi_<(\omega),\nonumber\\
\phi_>(\tau)&=\int_{\Lambda/b<|\omega|<\Lambda}\frac{d\omega}{2\pi}e^{-i\omega\tau}\phi_>(\omega),\nonumber
\end{align}
and likewise for $\b{S}_<(\tau)$ and $\b{S}_>(\tau)$, where $\Lambda\sim v_F/\xi$ is a high-energy cutoff and $b=1+d\ell$ is the rescaling parameter. Note that $\phi(\tau)=\phi_<(\tau)+\phi_>(\tau)$ and $\b{S}(\tau)=\b{S}_<(\tau)+\b{S}_>(\tau)$. To simplify the notation, we use the collective variable $\Phi\equiv(\phi,\b{S})$ to denote all the fields in the functional integral. Because $S_0$ is quadratic in $\Phi$, we have
\begin{align}
S[\Phi]=S_0[\Phi_<]+S_0[\Phi_>]+S_\perp[\Phi_<+\Phi_>].\nonumber
\end{align}
We define the effective action $S^<[\Phi_<]$ with reduced cutoff $\Lambda/b$ to be
\begin{align}
e^{-S^<[\Phi_<]}&\equiv\int\mathcal{D}\Phi_> e^{-S[\Phi]}\nonumber\\
&=e^{-S_0[\Phi_<]}Z_0^>
\left\langle e^{-S_\perp[\Phi_<+\Phi_>]}\right\rangle_>,\nonumber
\end{align}
where $Z_0^>=\int\mathcal{D}\Phi_> e^{-S_0[\Phi_>]}$ and $\langle\cdots\rangle_>$ denotes an expectation value with respect to $S_0[\Phi_>]$. Using the linked cluster theorem, we obtain
\begin{align}\label{SingleImpurityLinkedCluster}
S^<[\Phi_<]&=S_0[\Phi_<]+\left\langle S_\perp[\Phi_<+\Phi_>]\right\rangle_>\nonumber\\
&-\half\left(\left\langle S_\perp[\Phi_<+\Phi_>]^2\right\rangle_>-\left\langle S_\perp[\Phi_<+\Phi_>]\right\rangle_>^2\right),
\end{align}
to $\c{O}(J_\perp^2)$. The first order term contains the expectation value $\left\langle(S_<^\pm+S_>^\pm)e^{\pm i2\sqrt{\pi}(\phi_<+\phi_>)}\right\rangle_>$ which is given by the sum of two terms,
\begin{align}\label{SingleImpFirstOrderTerm1}
&S_<^\pm e^{\pm i2\sqrt{\pi}\phi_<}\left\langle e^{\pm i2\sqrt{\pi}\phi_>}\right\rangle_>
+e^{\pm i2\sqrt{\pi}\phi_<}\left\langle S_>^\pm e^{\pm i2\sqrt{\pi}\phi_>}\right\rangle_>\nonumber\\
&=S_<^\pm e^{\pm i2\sqrt{\pi}\phi_<}.
\end{align}
Using the unitary transformation $U$ introduced earlier enables us to compute correlators in the $\langle\cdots\rangle_>$ ensemble exactly, without having to expand in powers of $J_z$. Passing to the operator formalism, we consider $U=e^{i\lambda\phi_> S_>^z}$ which leaves $S^z_>$ and the vertex operators $e^{\pm i2\sqrt{\pi}\phi_>}$ unchanged, but under which the $S_>^\pm$ operators transform as $S_>^\pm\rightarrow US_>^\pm U^\dag=S_>^\pm e^{\pm i\lambda\phi_>}$ [Eq.~(\ref{SingleImpSpmTransform})]. Instead of choosing $\lambda=-2\sqrt{\pi}$ as in the decoupling limit, here we choose $\lambda=-J_za/\sqrt{\pi}v_FK$ which implies that the $J_z$ term in the transformed Hamiltonian cancels altogether.\cite{maciejko2009} Note that this cancellation occurs for any values of $J_z$ and $K$, i.e., we are \emph{not} assuming the decoupling limit Eq.~(\ref{SingleImpDecouplingLimit}). Performing this transformation, we trade correlators of $S_>^\pm$ and $e^{\pm i2\sqrt{\pi}\phi_>}$ in an ensemble governed by $S_\mathrm{TL}+S_z$ for correlators of $S_>^\pm e^{\pm i\lambda\phi_>}$ and $e^{\pm i2\sqrt{\pi}\phi_>}$, respectively, in an ensemble governed by $S_\mathrm{TL}$ alone. Applying these results to Eq.~(\ref{SingleImpFirstOrderTerm1}), we obtain
\begin{align}
\left\langle e^{\pm i2\sqrt{\pi}\phi_>}\right\rangle_>&=\left\langle e^{\pm i2\sqrt{\pi}\phi_>}\right\rangle_{>,\mathrm{TL}}=1,\nonumber\\
\left\langle S_>^\pm e^{\pm i2\sqrt{\pi}\phi_>}\right\rangle_>&=\left\langle S_>^\pm e^{\pm i2\sqrt{\pi}\chi\phi_>}\right\rangle_{>,\mathrm{TL}}\nonumber\\
&=\left\langle S_>^\pm\right\rangle_{>,\mathrm{TL}}\left\langle e^{\pm i2\sqrt{\pi}\chi\phi_>}\right\rangle_{>,\mathrm{TL}}=0,\nonumber
\end{align}
from which Eq.~(\ref{SingleImpFirstOrderTerm1}) follows. We have defined $\chi\equiv 1+\lambda/2\sqrt{\pi}=1-\rho J_z/2K$, which can be seen as a measure of the deviation from the decoupling limit $\chi=0$. As a result, we obtain
\begin{align}\label{SingleImpurityFirstOrder}
\left\langle S_\perp[\Phi_<+\Phi_>]\right\rangle_>=S_\perp[\Phi_<].
\end{align}
We now consider the second order term in Eq.~(\ref{SingleImpurityLinkedCluster}). The first term is
\begin{align}
&\left\langle S_\perp[\Phi_<+\Phi_>]^2\right\rangle_>=\left(\frac{J_\perp a}{2\pi\xi}\right)^2\int_0^\infty d\tau\int_0^\infty d\tau'\nonumber\\
&\hspace{5mm}\times\left\langle\left([S_<^+(\tau)+S_>^+(\tau)]e^{i2\sqrt{\pi}[\phi_<(\tau)+\phi_>(\tau)]}+\mathrm{c.c.}\right)\right.\nonumber\\
&\hspace{5mm}\times\left.\left([S_<^+(\tau')+S_>^+(\tau')]e^{i2\sqrt{\pi}[\phi_<(\tau')+\phi_>(\tau')]}+\mathrm{c.c.}\right)\right\rangle_>.\nonumber
\end{align}
Expressions containing the $\phi_<$ and $\b{S}_<$ fields factor out of the average. However, because the average $\langle\cdots\rangle_>$ is time-ordered, products of the noncommuting operators $S^+_<$ and $S^-_<$ need to be time-ordered as well. Using the unitary transformation introduced earlier, we obtain
\begin{align}
T_\tau S^\pm_<(\tau)S^\mp_<(\tau')=\left(\half+S^z_<\sgn(\tau-\tau')\right)e^{i\lambda[\phi_<(\tau)-\phi_<(\tau')]},\nonumber
\end{align}
where $T_\tau$ is the time-ordering operator in imaginary time, and the time dependence of $\phi_<$ and $S_<^z$ on the right-hand side of the equality is governed by $S_\mathrm{TL}[\Phi_<]$. The products $T_\tau S_<^\pm(\tau)S_<^\pm(\tau')$ are zero. The remaining expectation values are
\begin{align}\label{SingleImpAverage1}
\left\langle e^{\pm i2\sqrt{\pi}[\phi_>(\tau)-\phi_>(\tau')]}\right\rangle_>
&=\left\langle e^{\pm i2\sqrt{\pi}[\phi_>(\tau)-\phi_>(\tau')]}\right\rangle_{>,\mathrm{TL}}\nonumber\\
&=e^{4\pi D_>(\tau-\tau')},
\end{align}
and
\begin{align}\label{SingleImpAverage2}
&\left\langle S_>^\pm(\tau)S_>^\mp(\tau')e^{\pm i2\sqrt{\pi}[\phi_>(\tau)-\phi_>(\tau')]}\right\rangle_>\nonumber\\
&\hspace{28mm}=\left\langle S_>^\pm(\tau)S_>^\mp(\tau')\right\rangle_{>,\mathrm{TL}}e^{4\pi\chi^2 D_>(\tau-\tau')},\nonumber\\
&\hspace{28mm}=\half e^{4\pi\chi^2 D_>(\tau-\tau')},
\end{align}
where $D_>(\tau-\tau')\equiv\left\langle\phi_>(\tau)\phi_>(\tau')\right\rangle_{>,\mathrm{TL}}$ is the propagator associated to $S_\mathrm{TL}[\Phi_>]$. Correlators involving $S_<^\pm(\tau)S_<^\pm(\tau')$ vanish. Substituting Eq.~(\ref{SingleImpAverage1}) and (\ref{SingleImpAverage2}) in the expression for $\langle S_\perp^2\rangle_>$, we find three terms, one involving $\sin 2\sqrt{\pi}\chi(\phi_<-\phi_<')$, one involving $\cos 2\sqrt{\pi}\chi(\phi_<-\phi_<')$, and one involving $\cos 2\sqrt{\pi}(\phi_<-\phi_<')$, where we have denoted $\phi_<\equiv\phi_<(\tau)$ and $\phi_<'\equiv\phi_<(\tau')$. The fields $\phi_<$ are slow fields and their expectation value $\langle\phi_<\rangle$ vanishes in the unperturbed ensemble $S_\mathrm{TL}[\Phi_<]$ which governs their time dependence. Therefore, we will perform a gradient expansion of the sines and cosines in powers of time derivatives of $\phi_<$. The lowest order term for the sine is a single power of $\partial_\tau\phi_<\sim\omega\phi(\omega)$, which is a marginal operator at the Tomonaga-Luttinger fixed point. This term will lead to a renormalization of $J_z$, which is precisely what we are looking for. The lowest order term for the cosine is the quadratic term $\sim(\partial_\tau\phi_<)^2\sim\omega^2|\phi_<(\omega)|^2$, i.e., a quadratic kinetic energy term. However, the kinetic energy term Eq.~(\ref{SingleImpSTL}) in the unperturbed action is proportional to $|\omega|$, hence the $\omega^2$ term is irrelevant and can be neglected. In particular, this means that the Luttinger parameter $K$ is not renormalized. This is physically intuitive: a perturbation which exists only at $x=0$ cannot renormalize a bulk parameter.\cite{Giamarchi2003} Therefore, only the term containing $\sin 2\sqrt{\pi}\chi(\phi_<-\phi_<')$ needs to be kept.

The boson propagator $D_>(\tau-\tau')$ in Eq.~(\ref{SingleImpAverage1}) is given by
\begin{align}
D_>(\tau-\tau')
=\int_{\Lambda/b<|\omega|<\Lambda}\frac{d\omega}{2\pi}e^{-i\omega\tau}D(\omega)\nonumber,
\end{align}
where the propagator $D(\omega)=K/2|\omega|$ is read off of the unperturbed action Eq.~(\ref{SingleImpSTL}). We have
\begin{align}\label{SingleImpPropagFast}
D_>(\tau-\tau')&=K\int_{\Lambda/b}^\Lambda\frac{d\omega}{2\pi}\frac{1}{\omega}\cos\omega(\tau-\tau')\nonumber\\
&=\frac{K}{2\pi}\left.\left(\int\frac{d\omega}{\omega}\cos\omega(\tau-\tau')\right)\right|_{\Lambda/b}^\Lambda\nonumber\\
&=\frac{K}{2\pi}d\ell\cos\Lambda(\tau-\tau')+\c{O}(d\ell^2),
\end{align}
hence we obtain to $\c{O}(d\ell)$,
\begin{align}
&\half\left(\left\langle S_\perp^2\right\rangle_>-\left\langle S_\perp\right\rangle_>^2\right)\nonumber\\
&\hspace{10mm}=\left(\frac{J_\perp a}{2\pi\xi}\right)^2\int_0^\infty d\tau\int_0^\infty d\tau'
\left(e^{2Kd\ell\cos\Lambda(\tau-\tau')}-1\right)\nonumber\\
&\hspace{18mm}\times iS^z_<\sgn(\tau-\tau')\sin2\sqrt{\pi}\chi[\phi_<(\tau)-\phi_<(\tau')].\nonumber
\end{align}
We expand the exponential to $\c{O}(d\ell)$ and are led to consider the following integral,
\begin{align}
I[\phi_<]&=\int_0^\infty d\tau\int_0^\infty d\tau'\cos\Lambda(\tau-\tau')\nonumber\\
&\hspace{15mm}\times\sin2\sqrt{\pi}\chi[\phi_<(\tau)-\phi_<(\tau')]\sgn(\tau-\tau').\nonumber
\end{align}
Changing variables from $\tau$ and $\tau'$ to center-of-mass $T=\frac{1}{2}(\tau+\tau')$ and relative $t=\tau-\tau'$ variables, we observe that $\cos\Lambda(\tau-\tau')=\cos\Lambda t$ oscillates with a high frequency $\Lambda$. As a result, the integral over $t$ will be cut off at the short time $\sim\Lambda^{-1}$ and we can expand the functions of $\phi_<(\tau)-\phi_<(\tau')$ in powers of $t$. We obtain
\begin{align}
I[\phi_<]&\simeq\int_0^\infty dT\int_{-1/\Lambda}^{1/\Lambda}dt\left(2\sqrt{\pi}\chi|t|\partial_T\phi_<(T)+\c{O}(t^3)\right)\nonumber\\
&=2\sqrt{\pi}\chi\Lambda^{-2}\int_0^\infty dT\,\partial_T\phi_<+\c{O}(\Lambda^{-4}),\nonumber
\end{align}
and the effective action Eq.~(\ref{SingleImpurityLinkedCluster}) becomes
\begin{align}
S^<[\Phi_<]=S[\Phi_<]-\frac{\delta(J_za)}{\sqrt{\pi}v_F}\int_0^\infty d\tau S^zi\partial_\tau\phi_<,\nonumber
\end{align}
where the correction to $J_z$ is given by
\begin{align}
\frac{\delta(J_za)}{\pi v_F}=\left(\frac{J_\perp a}{\pi v_F}\right)^2K\chi d\ell,\nonumber
\end{align}
so that using the definition of $\chi$ in terms of $J_z$, the RG equation for $J_z$ is given by
\begin{align}\label{SingleImpRGJz}
\frac{dJ_z}{d\ell}=\rho K\left(1-\frac{\rho J_z}{2K}\right)J_\perp^2.
\end{align}
We have effectively computed the one-loop contribution to the RG equations. The tree level contributions are obtained by performing a scale transformation to restore the cutoff $\Lambda/b$ to its original value $\Lambda$, or alternatively\cite{CardyBook} by computing the scaling dimensions of the perturbation. In our case, we treated $S_0=S_\mathrm{TL}+S_z$ as the unperturbed action, hence the only perturbation is $S_\perp$. The scaling dimension $\Delta$ of an operator $O(\tau)$ is defined by\cite{CardyBook}
\begin{align}
\langle O(\tau)O^\dag(\tau')\rangle\sim|\tau-\tau'|^{-2\Delta},\nonumber
\end{align}
where the expectation value is taken at the appropriate fixed point, which in our case is the unperturbed ensemble governed by $S_0=S_\mathrm{TL}+S_z$. Defining $O_\perp\equiv S^+e^{i2\sqrt{\pi}\phi}+\mathrm{H.c.}$ and using the unitary transformation mentioned earlier, we have
\begin{align}
\langle O_\perp(\tau)O^\dag_\perp(\tau')\rangle_0&\sim\left\langle S^+(\tau)S^-(\tau')e^{i2\sqrt{\pi}\chi[\phi(\tau)-\phi(\tau')]}\right\rangle_{\mathrm{TL}}\nonumber\\
&\hspace{5mm}+\mathrm{c.c.}\nonumber\\
&\sim e^{4\pi\chi^2D(\tau-\tau')}=|\tau-\tau'|^{-2K\chi^2},\nonumber
\end{align}
where the full boson propagator, as opposed to the propagator of Eq.~(\ref{SingleImpPropagFast}) for the fast field $\phi_>$, is given by $D(\tau-\tau')=-(K/2\pi)\ln|\tau-\tau'|$. We therefore have $\Delta_\perp=K\chi^2$, and the RG equation\cite{CardyBook} for $J_\perp$ is given by
\begin{align}\label{SingleImpRGJperp}
\frac{dJ_\perp}{d\ell}=\left[1-K\left(1-\frac{\rho J_z}{2K}\right)^2\right]J_\perp.
\end{align}
Equations (\ref{SingleImpRGJz}) and (\ref{SingleImpRGJperp}) are the main result of this section, and are perturbative in $J_\perp$ but exact in $J_z$ and in $K$. In the weak coupling limit $\rho J_z\ll 1$, Eq.~(\ref{SingleImpRGJperp}) reduces to the poor man's scaling result Eq.~(\ref{SingleImpRGPerturbative1}), but Eq.~(\ref{SingleImpRGJz}) becomes
\begin{align}
\frac{dJ_z}{d\ell}=\rho KJ_\perp^2,\nonumber
\end{align}
which agrees with Eq.~(\ref{SingleImpRGPerturbative2}) except for a factor of $K$. Since $K\rightarrow 1$ in the noninteracting limit, we conclude that Eq.~(\ref{SingleImpRGPerturbative2}) is perturbative in the strength of electron-electron interactions in the helical liquid.

\begin{figure}[t]
\begin{center}
\includegraphics[width=3.5in,angle=0]{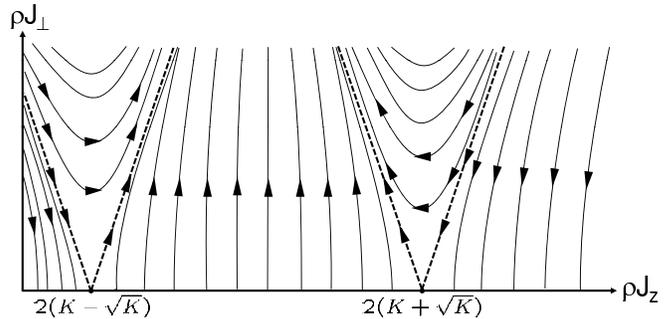}
\end{center}
\caption{Kosterlitz-Thouless RG flow for the single-impurity problem, with two critical points at $\rho J_z^{*,\pm}=2(K\pm\sqrt{K})$. The dotted lines correspond to the Kosterlitz-Thouless separatrix.}
\label{fig:RGFlowSingleImp}
\end{figure}

From Eq.~(\ref{SingleImpRGJz}) and (\ref{SingleImpRGJperp}) we immediately see that there is a quantum critical point at $J_\perp^*=0$ and $J_z^*$ such that $K\chi^2=1$. For a generic value of $K$ with $0<K<1$ there are in fact two critical points at $\rho J_z^*=2(K\pm\sqrt{K})$, which merge as $K\rightarrow 0$. In contrast, the RG equations (\ref{SingleImpRGPerturbative1}) and (\ref{SingleImpRGPerturbative2}) predict a single critical point at $\rho J_z^*=K-1=-g$ where we have defined $g\equiv 1-K$ as the strength of electron-electron interactions in the helical liquid. In the limit $g\ll 1$ of weak interactions, we find
\begin{align}
\rho J_z^{*,-}&\equiv 2(K-\sqrt{K})=-g+\c{O}(g^2),\label{SingleImpQCP1}\\
\rho J_z^{*,+}&\equiv 2(K+\sqrt{K})=4-3g+\c{O}(g^2).\label{SingleImpQCP2}
\end{align}
In other words, in the weak interaction limit $K\simeq 1$ we recover the critical point (\ref{SingleImpQCP1}) which has been previously predicted,\cite{Wu2006,maciejko2009} but we also find a new critical point [Eq.~(\ref{SingleImpQCP2})] at large $J_z$ which has been missed in previous studies. The RG flow in the $(J_z,J_\perp)$ plane for all $J_z$ and small $J_\perp$ and for different values of $K$ is easily obtained by a numerical solution of Eq.~(\ref{SingleImpRGJz}) and (\ref{SingleImpRGJperp}), and is of the Kosterlitz-Thouless type [Fig.~(\ref{fig:RGFlowSingleImp})] as could be expected for a single-channel Kondo impurity problem.\cite{anderson1970} For $J_z$ close to the $J_\perp=0$ critical points $J_z^*$, the Kosterlitz-Thouless separatrix is given by
\begin{align}
J_\perp^*=\pm\frac{1}{\sqrt{K}}(J_z-J_z^*)+\c{O}((J_z-J_z^*)^2),\nonumber
\end{align}
and separates a strong coupling antiferromagnetic (AF) phase where $J_\perp$ flows to infinity from a weak coupling ``local moment'' (LM) phase where $J_\perp$ flows to zero. The AF phase is a Kondo screened phase for which a qualitative description \`{a} la Nozi\`{e}res\cite{Nozieres1974} has been given previously.\cite{Wu2006,maciejko2009} The LM phase is described by an effective Hamiltonian of the form of Eq.~(\ref{HSingleImp}) but with $J_\perp=0$ and $J_z=J_z^\star$ where $J_z^\star$ (not to be confused with the critical point $J_z^*$) is the renormalized value of $J_z$, i.e., the $\ell\rightarrow\infty$ limit of the solution $J_z(\ell)$ of the RG equations (\ref{SingleImpRGJz}) and (\ref{SingleImpRGJperp}) which is finite in the LM phase [Fig.~(\ref{fig:RGFlowSingleImp})]. The spin-spin correlations are easily obtained from this effective Hamiltonian by applying the unitary transformation $U$ with $\lambda=-J_z^\star a/\sqrt{\pi}v_FK$ to remove the $S^z$ term from the effective Hamiltonian. We thus obtain in the LM phase
\begin{align}
\chi''_\perp(\omega)&=\frac{\pi\Lambda^{-\rho J_z^\star/2K}}{2\Gamma(\rho J_z^\star/2K)}|\omega|^{\rho J_z^\star/2K-1}\sgn\omega,\nonumber\\
\chi''_z(\omega)&=0.\nonumber
\end{align}

\subsection{Two-particle backscattering}\label{sec:SingleImp2PB}

The Hamiltonian (\ref{HSingleImp}) describes a spin-$\half$ local moment interacting by magnetic exchange with the helical liquid. However, as argued in the introduction, a generic quantum impurity on the edge of a QSH insulator can also give rise to a local 2-particle backscattering term, which is allowed by the topology of the QSH state.\cite{Wu2006,Xu2006} In the bosonized language, this amounts to adding to Eq.~(\ref{HSingleImp}) the term
\begin{align}\label{H2PBSingleImp}
H_\textrm{2PB}=\frac{\lambda_2 a}{\pi\xi}\cos 4\sqrt{\pi}\phi(0).
\end{align}
This term breaks the full $U(1)_s=\{e^{i\alpha Q_s}:\alpha\in[0,2\pi)\}$ symmetry of the original Hamiltonian (\ref{HSingleImp}) down to the subgroup $\{1,e^{i\pi Q_s}\}\cong\mathbb{Z}_2$. Indeed, this operator flips the spins of two conduction electrons and thus violates the conservation of the total $S^z$. However, this operator is allowed because it does not violate time-reversal symmetry. Recently, it has been realized that the inelastic backscattering term $(\partial_x^2\vartheta(0))e^{i2\sqrt{\pi}\phi(0)}+\mathrm{h.c.}$ with $\vartheta(x)=\int_{-\infty}^xdx'~\Pi(x')$ the dual boson, which is a conformal descendant of the time-reversal symmetry breaking single-particle backscattering operator $\cos 2\sqrt{\pi}\phi(0)$, does not itself break time-reversal symmetry and is thus an allowed perturbation.\cite{schmidt2011,lezmy2012} However, this operator has the scaling dimension $K+2$ which is always greater than one for repulsive interactions $0<K<1$, and is thus always irrelevant. Since we are only interested in the zero temperature phase diagram, this operator can be safely ignored here.

We first consider the effect of Eq.~(\ref{H2PBSingleImp}) in the decoupling limit $\rho J_z=2K$. Because any function of $\phi$ alone commutes with the unitary transformation $U$ [Eq.~(\ref{SingleImpPhiTransform})], the transformed Hamiltonian $\tilde{H}$ still exhibits the decoupling of the impurity spin from the conduction electrons,
\begin{align}
\tilde{H}=H_c[\phi,\Pi]+\frac{J_\perp a}{\pi\xi}\b{S}\cdot\hat{\b{e}},
\end{align}
but the conduction electron part of $\tilde{H}$ is
\begin{align}\label{BSG}
H_c[\phi,\Pi]=H_\mathrm{TL}[\phi,\Pi]+\frac{\lambda_2 a}{\pi\xi}\cos 4\sqrt{\pi}\phi(0),
\end{align}
i.e., the boundary sine-Gordon model. We still have $\langle\b{S}\cdot\hat{\b{e}}(\phi(0))\rangle_H=-\frac{1}{2}\sgn J_\perp$ with the unit vector $\hat{\b{e}}(\phi(0))$ defined in Eq.~(\ref{SingleImpUnitVectorPhi}), but now there is the possibility that the local CDW phase $2\sqrt{\pi}\phi(0)$ might get pinned because of the cosine potential in Eq.~(\ref{BSG}). As can be inferred from the RG equation\cite{Wu2006}
\begin{align}\label{RGlambda2SingleImp}
\frac{d\lambda_2}{d\ell}=(1-4K)\lambda_2,
\end{align}
this occurs when $K<1/4$, and $2\sqrt{\pi}\phi(0)$ is pinned in the ground state at $(n+\half)\pi$ for $\lambda_2>0$ and $n\pi$ for $\lambda_2<0$, with $n\in\mathbb{Z}$. From Eq.~(\ref{SingleImpUnitVectorPhi}) this means that $\hat{\b{e}}(\phi(0))=\pm\hat{\b{y}}$ for $\lambda_2>0$ and $\hat{\b{e}}(\phi(0))=\pm\hat{\b{x}}$ for $\lambda_2<0$, hence
\begin{align}
\langle S_y\rangle&=\pm\half\sgn J_\perp,\hspace{5mm}\lambda_2>0,\nonumber\\
\langle S_x\rangle&=\pm\half\sgn J_\perp,\hspace{5mm}\lambda_2<0,\nonumber
\end{align}
where the sign in $\pm$ is picked by spontaneous breaking of the $\mathbb{Z}_2$ symmetry. That spontaneous symmetry breaking is allowed in this $(0+1)$-dimensional problem at zero temperature can be seen by mapping the boundary sine-Gordon model (\ref{BSG}) to a 1D classical gas with long-ranged, logarithmic two-body interactions.\cite{Giamarchi2003} In contrast to the $\lambda_2=0$ case where the ground state is paramagnetic with $\langle\b{S}\rangle=0$, for $K<1/4$ and any $\lambda_2\neq 0$ the ground state is an Ising ferromagnet with $\langle\b{S}\rangle\neq 0$. For $K>1/4$, the 2-particle backscattering term is irrelevant and the ground state is paramagnetic.

%In the Ising ordered phase $K<1/4$, the transverse spin-spin correlation function is still given by $\chi_\perp(\tau)=\langle T_\tau S^+(\tau)S^-(0)\rangle_{\tilde{H}}\left\langle e^{i\lambda[\phi(\tau)-\phi(0)]}\right\rangle_{\tilde{H}}$, but the expectation value of the vertex operator is to be calculated for the boundary sine-Gordon model Eq.~(\ref{BSG}).

Away from the decoupling limit, one may wonder whether the scaling dimension of the 2-particle backscattering operator Eq.~(\ref{H2PBSingleImp}) deviates from its value $4K$ in the decoupling limit. Since the 2-particle backscattering operator commutes with the unitary transformation $U=e^{i\lambda\phi(0)S^z}$ with $\lambda=-J_za/\sqrt{\pi}v_FK$ for any value of $J_z$, its scaling dimension is independent of $J_z$ and is always equal to $4K$ for weak coupling $\rho\lambda_2,\rho J_\perp\ll 1$. Furthermore, in the single-impurity problem $K$ is a bulk property which is invariant under the $(0+1)$-dimensional RG flow.\cite{Giamarchi2003} Therefore, the RG equation (\ref{RGlambda2SingleImp}) is valid for all $J_z$, and the 2-particle backscattering term is relevant for $K<1/4$ and irrelevant for $K>1/4$ independent of $J_z$. More formally, we can repeat the perturbative analysis of Sec.~\ref{sec:SingleImpAYH} after adding the term $S_\mathrm{2PB}=\frac{\lambda_2a}{\pi\xi}\int_0^\infty d\tau\,\cos 4\sqrt{\pi}\phi$ to the action. The first order contribution is $\langle S_\mathrm{2PB}[\Phi_<+\Phi_>]\rangle_>=S_\mathrm{2PB}[\Phi_<]$ as expected. The second order contribution contains two terms, the mixed term $\langle S_\perp S_\mathrm{2PB}\rangle_>$ which vanishes and $\langle S_\mathrm{2PB}^2\rangle_>$ which only gives irrelevant terms. The only new contribution to the RG equations (\ref{SingleImpRGJz}) and (\ref{SingleImpRGJperp}) is the tree-level equation (\ref{RGlambda2SingleImp}).

\subsection{Phase diagram of the single-impurity problem}

\begin{figure}[t]
\begin{center}
\includegraphics[width=\columnwidth,angle=0]{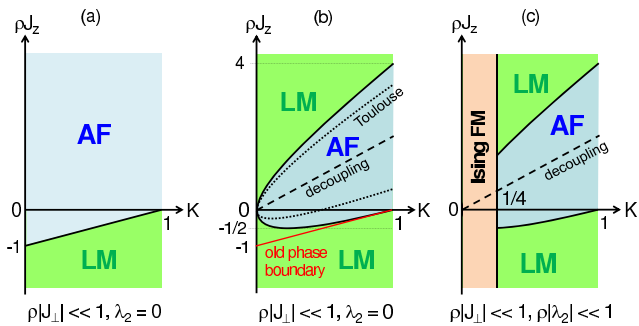}
\end{center}
\caption{Zero temperature phase diagram of a single Kondo impurity in the helical edge liquid of the quantum spin Hall state, for small $J_\perp$. (a) Previously derived (incorrect) phase diagram based on perturbation theory in $\rho J_z$ and $1-K$, with a Kondo screened strong coupling phase (AF) and an unscreened local moment phase (LM). (b) Phase diagram based on the RG equations (\ref{SingleImpRGJz}) and (\ref{SingleImpRGJperp}) exact in $\rho J_z$ and $K$. The decoupling limit $\rho J_z=2K$ (thicked dashed line) and the Toulouse limit (thick dotted line, Ref.~\onlinecite{tanaka2011}) both lie in the AF phase. The new AF-LM phase boundary (thick solid line) is contrasted with the previously derived phase boundary (thin solid line) of panel (a). (c) Phase diagram in the presence of impurity-induced 2-particle backscattering, with an additional symmetry-breaking Ising ferromagnetic phase (Ising FM).}
\label{fig:fig2}
\end{figure}

Based on the results just described, we can construct a revised zero temperature phase diagram for the single-impurity problem in the space of coupling constants $K$ and $J_z$ and in the limit $\rho|J_\perp|\ll 1$ and $\rho\lambda_2\ll 1$ (Fig.~\ref{fig:fig2}). The previously obtained poor man's scaling equations (\ref{SingleImpRGPerturbative1}) and (\ref{SingleImpRGPerturbative2}) predict the existence of a single Kondo screened strong coupling phase (AF) and a single unscreened local moment phase (LM) separated by a single phase boundary at $\rho J_z=K-1$ [Fig.~\ref{fig:fig2}(a)]. In particular, they predict that for antiferromagnetic $J_z>0$ a Kondo screened phase will always result. Our new RG equations (\ref{SingleImpRGJz}) and (\ref{SingleImpRGJperp}) predict a different topology for the phase diagram [Fig.~\ref{fig:fig2}(b)]. The AF phase is sandwiched between two LM phases. This leads to the surprising result that for large enough antiferromagnetic $J_z>0$, a LM phase will result, with no Kondo screening. The decoupling limit studied in Sec.~\ref{sec:SingleImpDecoupling} and the Toulouse limit studied in Ref.~\onlinecite{tanaka2011} both lie inside the AF phase. Fig.~\ref{fig:fig2}(a) and (b) both correspond to the absence of 2-particle backscattering $\lambda_2=0$. For $\lambda_2\neq 0$, the ground state is an Ising ferromagnet (Ising FM) for $K<1/4$ [Fig.~\ref{fig:fig2}(c)]. The phase diagrams are independent of the sign of $J_\perp$ as can be seen from the fact that the RG equations are symmetric under $J_\perp\rightarrow-J_\perp$.

\section{Kondo lattice problem}\label{sec:KondoLattice}

We now consider a regular array of spin-$\half$ magnetic impurities interacting with the helical liquid via exchange interactions. The Hamiltonian of this 1D Kondo lattice problem is given in the bosonized representation by
\begin{align}\label{HKL}
&H=H_\mathrm{TL}+H_z+H_\perp\nonumber\\
&\hspace{3mm}=\frac{v_F}{2}\int dx\left[K\Pi^2+\frac{1}{K}(\partial_x\phi)^2\right]-\frac{J_z a}{\sqrt{\pi}}\sum_rS^z_r\Pi(r)\nonumber\\
&\hspace{10mm}+\frac{J_\perp a}{2\pi\xi}\sum_r\left(S^+_re^{i2\sqrt{\pi}\phi(r)+i2k_Fr}+\mathrm{H.c.}\right),
\end{align}
where $\sum_r$ is a sum over the positions $r$ of the impurity spins $\b{S}_r$ which are equally separated by a distance $a$, i.e., $r=na$, $n\in\mathbb{Z}$. An important difference between the single-impurity Hamiltonian~(\ref{HSingleImp}) and the lattice Hamiltonian~(\ref{HKL}) is the presence of the $2k_Fr$ phase factors in the latter case, where $k_F$ is the Fermi wave vector of the helical liquid. Whereas the Fermi wave vector plays no role in the single-impurity problem, it plays an important role in the lattice problem, especially when 2-particle backscattering terms are considered (Sec.~\ref{sec:Lattice2PB}). As will be seen, the phase diagram of the Kondo lattice problem depends crucially on whether the system is at half-filling ($2k_Fa=\pi$) or away from half-filling ($2k_Fa\neq\pi$). Although this can be expected from the physics of the 1D Kondo lattice problem in an ordinary spinful 1D Fermi liquid,\cite{tsunetsugu1997} there are important differences which will be pointed out in due course. As in the single-impurity problem, the Hamiltonian (\ref{HKL}) has continuous $U(1)_c$ charge and $U(1)_s$ spin rotation symmetries, but the generator of $U(1)_s$ now contains the $z$ component of the total impurity spin, $Q_s=\half\int dx(-\frac{1}{\sqrt{\pi}})\Pi+\sum_rS^z_r$.

The model (\ref{HKL}) is similar to an orbital analog of the 1D Kondo lattice model studied earlier,\cite{emery1993} in which physical impurity spins were replaced by impurity pseudospins which couple to the orbital states of the conduction electrons rather than to their true spin. In both models, the $z$ component of the impurity (pseudo-)spin couples to the local electronic current $j_x(r)=-\frac{1}{\sqrt{\pi}}\Pi(r)$. In the model of Ref.~\onlinecite{emery1993}, this occurred because the impurity pseudospin was assumed to carry an electric dipole moment which would thus couple to the conduction electron current. In the helical liquid however, this is the form taken by the physical magnetic exchange interaction, because the local current of the conduction electrons corresponds to the $z$ component of their local magnetization, due to the helical property of the QSH edge states.

\subsection{Decoupling limit}\label{sec:LatticeDecouplingLimit}

The unitary transformation of Sec.~\ref{sec:SingleImpDecoupling} can be generalized to the lattice case\cite{emery1993} by choosing $U=\exp(i\lambda\sum_r\phi(r)S^z_r)$. For $\lambda=-2\sqrt{\pi}$ and in the decoupling limit $\rho J_z=2K$, the transformed Hamiltonian $\tilde{H}=UHU^\dag$ reads
\begin{align}\label{KLTransformedH}
\tilde{H}=H_\mathrm{TL}[\phi,\Pi]+\frac{J_\perp a}{\pi\xi}\sum_r\b{S}_r\cdot\hat{\b{e}}_r,
\end{align}
where $\hat{\b{e}}_r=\hat{\b{x}}\cos 2k_Fr-\hat{\b{y}}\sin 2k_Fr$. As in the single-impurity case, the impurity spins decouple from the conduction electrons: the two terms in Eq.~(\ref{KLTransformedH}) commute and can be diagonalized independently. Furthermore, we have $[\b{S}_r\cdot\hat{\b{e}}_r,\tilde{H}]=0$ for all impurity sites $r$, hence $\b{S}_r\cdot\hat{\b{e}}_r$ is a good quantum number for all $r$ under $\tilde{H}$. In the ground state of $\tilde{H}$, we have $\b{S}_r\cdot\hat{\b{e}}_r=-\half\sgn J_\perp$ for all $r$, which appears to indicate that the system has long-range helical spin-density-wave (SDW) order in the $xy$ plane. However, this is not necessarily so for the same reason as before: $\b{S}_r\cdot\hat{\b{e}}_r$ is not a good quantum number under the original Hamiltonian $H$. We have
\begin{align}
-\half\sgn J_\perp=\langle\b{S}_r\cdot\hat{\b{e}}_r\rangle_{\tilde{H}}&=\langle U^\dag\b{S}_rU\cdot\hat{\b{e}}_r\rangle_H\nonumber\\
&=\langle\b{S}_r\cdot\hat{\b{e}}_r(\phi(r))\rangle_H,\nonumber
\end{align}
where we define the unit vector
\begin{align}\label{LatticeCDWUnitVector}
\hat{\b{e}}_r(\phi(r))=\hat{\b{x}}\cos\alpha_r-\hat{\b{y}}\sin\alpha_r,
\end{align}
with $\alpha_r\equiv 2\sqrt{\pi}\phi(r)+2k_Fr$. In this case the orientation of each impurity spin in the $xy$ plane is controlled both by the local dynamical phase $2k_Fr$ and the local CDW phase $2\sqrt{\pi}\phi(r)$ of the helical conduction electrons. If $\phi$ fluctuates strongly enough as is the case at the Tomonaga-Luttinger fixed point of the conduction electrons, there is no long-range helical spin-density-wave (SDW) order. The impurity spin part of the ground state wave function for the transformed Hamiltonian (\ref{KLTransformedH}) is easily found,
\begin{align}\label{LocalCDWOrderLattice}
|0\rangle_{\tilde{H}}=\prod_r\frac{1}{\sqrt{2}}\left(
\begin{array}{c}
e^{i2k_Fr} \\
-\sgn J_\perp
\end{array}
\right),
\end{align}
in the $S^z_r$ direct product basis. As in the single-impurity case, the correlation functions of the impurity spins can be evaluated with the help of the unitary transformation $U$. Defining $\b{S}_r^\perp\equiv S^x_r\hat{\b{x}}+S^y_r\hat{\b{y}}$, we obtain
\begin{align}\label{LatticeChiPerpTime}
&\chi_\perp(r-r',\tau)=\langle T_\tau\b{S}^\perp_r(\tau)\cdot\b{S}^\perp_{r'}(0)\rangle_H\nonumber\\
&\hspace{10mm}=\half\Bigl(\langle T_\tau S_r^+(\tau)S_{r'}^-(0)\rangle_{\tilde{H}}\left\langle e^{i\lambda[\phi(r,\tau)-\phi(r',0)]}\right\rangle_{\tilde{H}}
\nonumber\\
&\hspace{19mm}+\mathrm{c.c.}\Bigr)
\nonumber\\
&\hspace{10mm}=\frac{1}{4}\frac{\cos 2k_F(r-r')+e^{-\omega_\perp|\tau|}\delta(r-r')}{\left[\left(\frac{r-r'}{\xi}\right)^2+(\Lambda|\tau|)^2\right]^K},
\end{align}
and
\begin{align}
\chi_z(r-r',\tau)&=\langle T_\tau S^z_r(\tau)S^z_{r'}(0)\rangle_H\nonumber\\
&=\langle T_\tau S^z_r(\tau)S^z_{r'}(0)\rangle_{\tilde{H}}\nonumber\\
&={\textstyle\frac{1}{4}}e^{-\omega_\perp|\tau|}\delta(r-r').
\end{align}
In the local limit $r-r'=0$, the correlation functions of the Kondo lattice reduce to those of the single-impurity problem Eq.~(\ref{SingleImpChiTransverseTime}) and Eq.~(\ref{SingleImpChizzTime}). We define the momentum- and frequency-dependent impurity spin transverse susceptibility $\chi''_\perp(q,\omega)$ as the imaginary part of the Fourier transform of the corresponding retarded correlation function $\chi^R_\perp(r-r',t)=i\theta(t)\langle[S^+_r(t),S^-_{r'}(0)]\rangle$. As can be seen from Eq.~(\ref{LatticeChiPerpTime}), the contribution of the first term in Eq.~(\ref{LatticeChiPerpTime}) is almost the same as the $2k_F$ part of the particle-hole susceptibility of the spinless 1D electron gas which was calculated by Luther and Peschel.\cite{luther1974b} We obtain
\begin{widetext}
\begin{align}\label{LatticeChiPerp}
&\chi''_\perp(q,\omega)=A(|\omega|-\omega_\perp)^{2K-1}\theta(|\omega|-\omega_\perp)\sgn\omega
+B\left[\left|\omega^2-v_F^2(q-2k_F)^2\right|^{K-1}\theta(|\omega|-v_F|q-2k_F|)\right.\nonumber\\
&\hspace{75mm}\left.+\left|\omega^2-v_F^2(q+2k_F)^2\right|^{K-1}\theta(|\omega|-v_F|q+2k_F|)\right]\sgn\omega,
\end{align}
\end{widetext}
%\begin{align}\label{LatticeChiPerp}
%&\chi''_\perp(q,\omega)=A(|\omega|-\omega_\perp)^{2K-1}\theta(|\omega|-\omega_\perp)\sgn\omega\nonumber\\
%&\hspace{3mm}+B\left[\left|\omega^2-v_F^2(q-2k_F)^2\right|^{K-1}\theta(|\omega|-v_F|q-2k_F|)\right.\nonumber\\
%&\hspace{8mm}\left.+\left|\omega^2-v_F^2(q+2k_F)^2\right|^{K-1}\theta(|\omega|-v_F|q+2k_F|)\right]\nonumber\\
%&\hspace{8mm}\times\sgn\omega,
%\end{align}
where $A$ and $B$ are $(q,\omega)$-independent constants. The longitudinal susceptibility is given by
\begin{align}\label{LatticeChiz}
\chi''_z(q,\omega)=\frac{\pi}{4}\delta(|\omega|-\omega_\perp)\sgn\omega,
\end{align}
that is, it is purely local ($q$-independent) and equal to the single-impurity susceptibility Eq.~(\ref{SingleImpChizz}).

\subsection{Away from the decoupling limit: 2D Coulomb gas approach}\label{sec:2DCG}

As in the single-impurity case, the results Eq.~(\ref{LatticeChiPerp}) and (\ref{LatticeChiz}) which we found for the impurity spin susceptibilities are valid only in the decoupling limit $\rho J_z=2K$. In this section, we derive RG equations which will allow us to explore the phase diagram of the Hamiltonian (\ref{HKL}) away from that special limit. One way to proceed is to take the continuum limit of the impurity lattice at the outset and bosonize it. One then obtains a problem of two coupled boson fields for which RG equations can be derived either directly or by first mapping it to a classical 2D Coulomb gas problem.\cite{Gogolin} Taking the continuum limit of the impurity lattice is usually done by first adding by hand to the Hamiltonian (\ref{HKL}) a short-range exchange interaction term of the form $\sim J_H\sum_r\b{S}_r\cdot\b{S}_{r+a}$, the XY part of which generates a standard Tomonaga-Luttinger kinetic term.\cite{Giamarchi2003} The resulting Hamiltonian is of the Kondo-Heisenberg form. On the other hand, if one were to integrate out the conduction electron field $\phi$ in Eq.~(\ref{HKL}), one would generate long-range RKKY-type spin-spin interaction terms\cite{gao2009} of the form $\sim\sum_{rr'}J_{rr'}^{\sigma\sigma'}S_r^\sigma S_{r'}^{\sigma'}$, with $J_{rr'}^{\sigma\sigma'}\propto|r-r'|^{-\gamma}$ for some power $\gamma$ which depends on the Luttinger parameter $K$ of the helical liquid. Therefore, it is not clear that the Kondo-Heisenberg model with finite $J_H$, or anisotropic versions thereof, faithfully represents the original Kondo lattice model (\ref{HKL}). For that reason, we follow the approach of Novais \emph{et al.}\cite{novais2002} which does not require the adding by hand of a kinetic term for the impurity spins. This approach is essentially an extension of the Anderson-Yuval-Hamann procedure to the lattice case, where the Kondo lattice problem (\ref{HKL}) is mapped to a classical 2D Coulomb gas, for which RG equations can be derived using the real-space renormalization procedure introduced by Kosterlitz.\cite{kosterlitz1974} The main steps of the procedure are as follows.\cite{novais2002} As in Sec.~\ref{sec:SingleImpAYH} for the single-impurity problem, we use the unitary transformation $U$ to eliminate the $J_z$ term in Eq.~(\ref{HKL}) and formally expand the partition function in powers of $J_\perp$.\cite{anderson1969,yuval1970} We then perform the path integral over impurity spins and over the conduction electron field $\phi$ in the Tomonaga-Luttinger ensemble. The resulting partition function is that of a classical 2D gas of particles with unit charge $m=\pm 1$ interacting through a two-body logarithmic potential. It is well-known that this problem is equivalent to the classical 2D XY model and that the associated RG equations are the Kosterlitz-Thouless equations.\cite{kosterlitz1973,kosterlitz1974}

The quantum partition function associated to Eq.~(\ref{HKL}) is
\begin{align}\label{Z}
Z=\int\mathcal{D}\phi\left(\prod_r\int\mathcal{D}\b{S}_r\right)e^{-S[\phi,\b{S}_r]}.
\end{align}
Because the partition function is invariant under unitary transformations, we can choose to evaluate Eq.~(\ref{Z}) using the transformed Hamiltonian $\tilde{H}=UHU^\dag$ with $\lambda=-J_za/\sqrt{\pi}v_FK$, in which case the $J_z$ term disappears and the Euclidean action $S$ in Eq.~(\ref{Z}) becomes
\begin{align}\label{LatticeEuclideanAction}
&S[\phi,\b{S}_r]=S_\mathrm{TL}+S_\perp+S_\mathrm{WZ}\nonumber\\
&\hspace{5mm}=\frac{v_F}{2K}\int_0^\infty d\tau\int dx\left[(\partial_x\phi)^2+\frac{1}{v_F^2}(\partial_\tau\phi)^2\right]\nonumber\\
&\hspace{5mm}+\frac{J_\perp a}{2\pi\xi}\int_0^\infty d\tau\sum_r\left(S^+_r(\tau)e^{i2\sqrt{\pi}\chi\phi(r,\tau)+i2k_Fr}+\mathrm{c.c.}\right)\nonumber\\
&\hspace{5mm}+\sum_rS_\mathrm{WZ}[\b{S}_r],
\end{align}
where $\chi=1-\rho J_z/2K$ as in Sec.~\ref{sec:SingleImpAYH}, and $S_\mathrm{WZ}[\b{S}_r]$ in the last line of Eq.~(\ref{LatticeEuclideanAction}) is the Wess-Zumino term for a single impurity spin. Here again we do not require the explicit form of the Wess-Zumino term because spin-spin correlators will be evaluated using the operator formalism. We formally expand the partition function Eq.~(\ref{Z}) in powers of $J_\perp$,
\begin{align}\label{ZTaylorExpand}
Z=&\int\mathcal{D}\phi\left(\prod_r\int\mathcal{D}\b{S}_r\right)e^{-\left(S_\mathrm{TL}[\phi]+\sum_rS_\mathrm{WZ}[\b{S}_r]\right)}\nonumber\\
&\times
\sum_{N=0}^\infty\left(\frac{J_\perp a}{2\pi\xi}\right)^{N}\frac{1}{N!}\left(\prod_{\ell=1}^{N}\int_0^\infty d\tau_\ell\sum_{r_\ell}\right)
\nonumber\\
&\times\left(S^+_{r_1}(\tau_1)e^{i2\sqrt{\pi}\chi\phi(r_1,\tau_1)+i2k_Fr_1}+\mathrm{c.c.}\right)\nonumber\\
&\times\cdots
\left(S^+_{r_{N}}(\tau_{N})e^{i2\sqrt{\pi}\chi\phi(r_{N},\tau_{N})+i2k_Fr_{N}}+\mathrm{c.c.}\right).
\end{align}
The factors involving impurity spin flips can be written as
\begin{align}\label{spinflipIsing}
&S^+_{r_\ell}(\tau_\ell)e^{i2\sqrt{\pi}\chi\phi(r_\ell,\tau_\ell)+i2k_Fr_\ell}+\mathrm{c.c.}\nonumber\\
&\hspace{10mm}=\sum_{m_\ell=\pm 1}S^{m_\ell}_{r_\ell}(\tau_\ell)e^{[i2\sqrt{\pi}\chi\phi(r_\ell,\tau_\ell)+i2k_Fr_\ell]m_\ell},
\end{align}
with the obvious symbolic notation $S^m\equiv S^\pm$ for $m=\pm 1$, where we have introduced $N$ Ising variables $m_\ell=\pm 1$, $\ell=1,\ldots,N$ which correspond to spin flip events in spacetime. Substituting Eq.~(\ref{spinflipIsing}) in Eq.~(\ref{ZTaylorExpand}), we obtain
\begin{align}\label{ZIsingvariables}
Z&=\sum_{N=0}^\infty\frac{1}{N!}\left(\frac{J_\perp a}{2\pi\xi}\right)^{N}
\sum_{\{m\}}\int\mathcal{D}\boldsymbol{\eta}
\int\mathcal{D}\phi\left(\prod_r\int\mathcal{D}\b{S}_r\right)\nonumber\\
&\times e^{-S_\mathrm{TL}[\phi]}
\exp\left\{\sum_{\ell=1}^{N}\left[i2\sqrt{\pi}\chi\phi(\boldsymbol{\eta}_\ell)m_\ell+i2k_Fr_\ell m_\ell\right]\right\}\nonumber\\
&\times e^{-\sum_rS_\mathrm{WZ}[\b{S}_r]}\prod_{\ell=1}^{N}S^{m_\ell}_{r_\ell}(\tau_\ell),
\end{align}
where we introduce the 2D coordinates $\boldsymbol{\eta}\equiv(r,\tau)$ and the associated integration measure
\begin{align}
\int\mathcal{D}\boldsymbol{\eta}=\prod_{\ell=1}^{N}\int d^2\boldsymbol{\eta}_\ell\equiv\prod_{\ell=1}^{N}\int_0^\infty d\tau_\ell\sum_{r_\ell},
\nonumber
\end{align}
and we denote the sum over all possible configurations of the Ising variables $m_\ell$ by
\begin{align}
\sum_{\{m\}}\equiv\prod_{\ell=1}^{N}\sum_{m_\ell=\pm 1}.\nonumber
\end{align}
The path integral over impurity spins in Eq.~(\ref{ZIsingvariables}) can be performed first,
\begin{align}\label{averageprodspins}
&\left(\prod_r\int\mathcal{D}\b{S}_r\right)e^{-\sum_rS_\mathrm{WZ}[\b{S}_r]}\prod_{\ell=1}^{N}S^{m_\ell}_{r_\ell}(\tau_\ell)\nonumber\\
&\hspace{10mm}=Z_S\left\langle T_\tau S_{r_1}^{m_1}(\tau_1)S_{r_2}^{m_2}(\tau_2)\cdots S_{r_{N}}^{m_{N}}(\tau_{N})\right\rangle,
\end{align}
where $Z_S\equiv\left(\prod_r\int\mathcal{D}\b{S}_r\right)e^{-\sum_rS_\mathrm{WZ}[\b{S}_r]}$ is the partition function of the unperturbed impurity spins. Since only the Wess-Zumino term appears in the action, the expectation value on the right-hand side of Eq.~(\ref{averageprodspins}) is with respect to a zero Hamiltonian. As a result, the spin operators have no time dependence. However, the order of the operators does still matter because of the time-ordering operator. We calculate the expectation value in the $S^z$ basis,
\begin{align}\label{averageprodspins2}
&Z_S\left\langle T_\tau S_{r_1}^{m_1}(\tau_1)S_{r_2}^{m_2}(\tau_2)\cdots S_{r_{N}}^{m_{N}}(\tau_{N})\right\rangle\nonumber\\
&\hspace{10mm}=\sum_{\{S^z\}}\left\langle\{S^z\}\left|S_{r_1}^{m_1}S_{r_2}^{m_2}\cdots S_{r_{N}}^{m_{N}}\right|\{S^z\}\right\rangle,
\end{align}
where we assume the ordering $\tau_1>\tau_2>\cdots>\tau_{N}$. The nonvanishing of the correlation function (\ref{averageprodspins}) imposes some constraints on the Ising variable configurations $\{m\}$. First, because only $S^+$ and $S^-$ operators appear in Eq.~(\ref{averageprodspins}) with no $S^z$ operators, $N$ must be even. Second, in order for the final state to be the same as the initial state, there must be an equal number of $S^+$ and $S^-$ operators, i.e., $\sum_{\ell=1}^{N}m_\ell=0$. This is a global neutrality condition which is typical of sine-Gordon and Coulomb gas models.\cite{nienhuis} However, there are two more constraints on $\{m\}$ which are specific to Kondo models. Since for $S=\half$ spins we have $(S^\pm_r)^2=0$ for each $r$, the Ising variable $m_\ell$ must necessarily alternate in imaginary time\cite{anderson1969,yuval1970} for fixed $r=r_\ell$. For the expectation value in Eq.~(\ref{averageprodspins2}) to be nonzero, this means that in addition to the global neutrality condition we have a ``local'' neutrality condition $\sum_{r_\ell=r}m_\ell=0$ for each $r$.\cite{novais2002} These constraints can be illustrated by examples with few spins. For $N=2$, we have
\begin{align}
&\sum_{\{S^z\}}\left\langle\{S^z\}\left|S_{r_1}^{m_1}S_{r_2}^{m_2}\right|\{S^z\}\right\rangle\nonumber\\
&\hspace{12mm}=\delta_{m_1+m_2,0}\delta_{r_1=r_2}\sum_{\{S^z\}}\left\langle\{S^z\}\left|S_{r_1}^{m_1}S_{r_1}^{-m_1}\right|\{S^z\}\right\rangle\nonumber\\
&\hspace{12mm}=\delta_{m_1+m_2,0}\delta_{r_1=r_2}\sum_{\{S^z\}}\delta_{S^z_{r_1}=\frac{1}{2}m_1}\nonumber\\
&\hspace{12mm}=2^{N_\mathrm{sites}-1}\delta_{m_1+m_2,0}\delta_{r_1=r_2},\nonumber
\end{align}
where $N_\mathrm{sites}$ is the number of impurity sites. For $N=4$, we have
\begin{align}
&\sum_{\{S^z\}}\left\langle\{S^z\}\left|S_{r_1}^{m_1}S_{r_2}^{m_2}S_{r_3}^{m_3}S_{r_4}^{m_4}\right|\{S^z\}\right\rangle\nonumber\\
&\hspace{5mm}=\delta_{\sum_{\ell=1}^4m_\ell,0}\nonumber\\
&\hspace{5mm}\times\sum_{\{S^z\}}
\left(\delta_{m_1+m_2,0}\delta_{r_1=r_2}\delta_{r_3=r_4}\delta_{S^z_{r_1}=\frac{1}{2}m_1}\delta_{S^z_{r_3}=\frac{1}{2}m_3}
\right.\nonumber\\
&\hspace{15mm}+\delta_{m_1+m_3,0}\delta_{r_1=r_3}\delta_{r_2=r_4}\delta_{S^z_{r_1}=\frac{1}{2}m_1}\delta_{S^z_{r_2}=\frac{1}{2}m_2}
\nonumber\\
&\left.\hspace{15mm}+\delta_{m_1+m_4,0}\delta_{r_1=r_4}\delta_{r_2=r_3}\delta_{S^z_{r_1}=\frac{1}{2}m_1}\delta_{S^z_{r_2}=\frac{1}{2}m_2}
\right)\nonumber\\
&\hspace{5mm}=2^{N_\mathrm{sites}-2}\delta_{\sum_{\ell=1}^4m_\ell,0}
\left(\delta_{m_1+m_2,0}\delta_{r_1=r_2}\delta_{r_3=r_4}
\right.\nonumber\\
&\hspace{36mm}+\delta_{m_1+m_3,0}\delta_{r_1=r_3}\delta_{r_2=r_4}
\nonumber\\
&\left.\hspace{36mm}+\delta_{m_1+m_4,0}\delta_{r_1=r_4}\delta_{r_2=r_3}\right).\nonumber
\end{align}
The Kronecker deltas enforce the local neutrality condition for each impurity site. The average contains a factor of $2^{-N/2}$ which can be absorbed in the factor containing $J_\perp$ [see Eq.~(\ref{ZIsingvariables})]. The factor of $2^{N_\mathrm{sites}}$ is simply the partition function $Z_S$ of the unperturbed impurity spins. The partition function Eq.~(\ref{ZIsingvariables}) becomes
\begin{align}\label{ZBeforeIntegrateOutBoson}
Z&=Z_S\sum_{N=0}^\infty\frac{1}{N!}\left(\frac{J_\perp a}{2\sqrt{2}\pi\xi}\right)^{N}
\sum_{\{m\}}^\prime\int\mathcal{D}\boldsymbol{\eta}\int\mathcal{D}\phi\,e^{-S_\mathrm{TL}[\phi]}\nonumber\\
&\hspace{5mm}\times\exp\left\{\sum_{\ell=1}^{N}\left[i2\sqrt{\pi}\chi\phi(\boldsymbol{\eta}_\ell)m_\ell+i2k_Fr_\ell m_\ell\right]\right\},
\end{align}
where the prime on the sum over $\{m\}$ indicates the neutrality constraints mentioned earlier. The local neutrality condition $\sum_{r_\ell=r}m_\ell=0$ for each $r$ implies that $\sum_\ell r_\ell m_\ell=0$, hence the term proportional to $2k_F$ in Eq.~(\ref{ZBeforeIntegrateOutBoson}) vanishes. The path integral over $\phi$ is Gaussian and can be performed exactly,
\begin{align}\label{bosonaverageZ}
&\int\mathcal{D}\phi\,e^{-S_\mathrm{TL}[\phi]}
\exp\left\{\sum_{\ell=1}^{N}\left[i2\sqrt{\pi}\chi\phi(\boldsymbol{\eta}_\ell)m_\ell+i2k_Fr_\ell m_\ell\right]\right\}\nonumber\\
&\hspace{7mm}=Z_\mathrm{TL}\exp\left\{-2\pi\chi^2\sum_{\ell\ell'}m_\ell m_{\ell'}\langle T_\tau\phi(\boldsymbol{\eta}_\ell)\phi(\boldsymbol{\eta}_{\ell'})\rangle_\mathrm{TL}\right\},
\end{align}
where $Z_\mathrm{TL}$ is the partition function of the Tomonaga-Luttinger liquid. We see from Eq.~(\ref{TomonagaLuttingerPropagator}) that $\langle T_\tau\phi(\boldsymbol{\eta}')\phi(\boldsymbol{\eta}')\rangle\rightarrow\ln(1)=0$ for $|\boldsymbol{\eta}-\boldsymbol{\eta}'|\ll\xi$, hence the $\ell=\ell'$ term can be removed from Eq.~(\ref{bosonaverageZ}). We have therefore rewritten the partition function of the Kondo lattice problem (\ref{HKL}) as that of a 2D classical Coulomb gas,\cite{nienhuis}
\begin{align}\label{ZCG}
Z=Z_\mathrm{TL}Z_S\sum_{N=0}^\infty\sum_{\{m\}}^\prime\frac{Y^N}{N!}\int_{|\Delta\b{x}|>\xi}\mathcal{D}\b{x}\, e^{-S_\mathrm{CG}},
\end{align}
where $Y\equiv\rho|J_\perp|/2\sqrt{2}$ is the fugacity of the gas, and
\begin{align}\label{SCGno2PB}
S_\mathrm{CG}=-\half g\sum_{\ell\neq\ell'}m_\ell m_{\ell'}\ln\frac{|\b{x}_\ell-\b{x}_{\ell'}|}{\xi},
\end{align}
is a two-body logarithmic interaction potential with interaction strength $g\equiv 2K\chi^2$. We have introduced 2D coordinates with dimensions of length $\b{x}=(r,v_F\tau)$. Approximating the discrete sum over impurity sites by an integral $\sum_r\simeq\int\frac{dr}{\xi}$, the integration measure is
\begin{align}\label{CoulombGasIntegrationMeasure}
\int\mathcal{D}\b{x}=\prod_{\ell=1}^{N}\int\frac{d^2\b{x}_\ell}{\xi^2}.
\end{align}
The subscript $|\Delta\b{x}|>\xi$ in Eq.~(\ref{ZCG}) signifies that the configuration space integral (\ref{CoulombGasIntegrationMeasure}) is subject to the hard-core constraint $|\b{x}_\ell-\b{x}_{\ell'}|>\xi$ for all $\ell\neq\ell'$. As mentioned previously, this constraint comes from the short-distance behavior of the Tomonaga-Luttinger propagator Eq.~(\ref{TomonagaLuttingerPropagator}). We also note that the dimensionless fugacity $Y$ does not depend on the sign of $J_\perp$, because $N$ is even. Finally, the unperturbed partition functions $Z_\mathrm{TL}$ and $Z_S$ in Eq.~(\ref{ZCG}) do not contain any thermodynamic singularities and will be ignored in what follows.

The RG equations for the single-component Coulomb gas (\ref{ZCG}) are the well-known Kosterlitz-Thouless equations,\cite{kosterlitz1973,kosterlitz1974}
\begin{align}
\frac{dY}{d\ell}=\left(2-\half g\right)Y,\label{KTeq1}\\
\frac{dg}{d\ell}=-2\pi^2Y^2g^2.\label{KTeq2}
\end{align}
There is a quantum phase transition in the 2DXY universality class at the position $g=g_c(Y)$ of the Kosterlitz-Thouless separatrix, which for small $Y$ is given by $g_c(Y)=4+8\pi Y+\c{O}(Y^2)$. For $\rho J_\perp\ll 1$ corresponding to $Y\ll 1$, the transition occurs on a curve in the $(K,J_z)$ plane defined by $K\chi^2=2$.

\subsection{Two-particle backscattering}\label{sec:Lattice2PB}

As discussed in Sec.~\ref{sec:SingleImp2PB}, a quantum impurity on the edge of a QSH insulator can generally give rise to 2-particle backscattering processes. In the case of a regular array of impurities, this corresponds to adding to the Kondo lattice Hamiltonian (\ref{HKL}) the term
\begin{align}\label{H2PBlattice}
H_\textrm{2PB}=\frac{\lambda_2a}{\pi\xi}\sum_r\cos[4\sqrt{\pi}\phi(r)+4k_Fr],
\end{align}
in the boson representation. As in the single-impurity case, $H_\textrm{2PB}$ explicitly breaks the continuous $U(1)_s$ spin rotation symmetry of Eq.~(\ref{HKL}) to a discrete $\mathbb{Z}_2$ symmetry $\{1,e^{i\pi Q_s}\}$ where $Q_s$ is given at the end of the paragraph following Eq.~(\ref{HKL}).

We first consider the effect of Eq.~(\ref{H2PBlattice}) in the decoupling limit $\rho J_z=2K$. Following the same steps as in Sec.~\ref{sec:LatticeDecouplingLimit}, the transformed Hamiltonian $\tilde{H}=UHU^\dag$ is
\begin{align}
\tilde{H}=H_c[\phi,\Pi]+\frac{J_\perp a}{\pi\xi}\sum_r\b{S}_r\cdot\hat{\b{e}}_r,\nonumber
\end{align}
where
\begin{align}\label{HcLatticeDecoupling}
H_c[\phi,\Pi]=H_\mathrm{TL}[\phi,\Pi]+\frac{\lambda_2 a}{\pi\xi}\sum_r\cos[4\sqrt{\pi}\phi(r)+4k_Fr],
\end{align}
i.e., the transformed Hamiltonian in the conduction electron sector is a periodic sine-Gordon model. This problem can be solved exactly in two limits: at the Luther-Emery point\cite{luther1974} $K=1/4$ where the problem can be mapped to free fermions, and at the free boson point $K\rightarrow 0$.

We first discuss the solution of Eq.~(\ref{HcLatticeDecoupling}) at the Luther-Emery point $K=1/4$. We rescale the boson fields $\Pi\rightarrow\tilde{\Pi}=\half\Pi$ and $\phi\rightarrow\tilde{\phi}=2\phi$ which preserves the canonical commutation relations $[\tilde{\phi}(x),\tilde{\Pi}(x')]=[\phi(x),\Pi(x')]$. The Hamiltonian (\ref{HcLatticeDecoupling}) becomes
\begin{align}\label{HLatticeLutherEmery}
\tilde{H}_c[\tilde{\phi},\tilde{\Pi}]=&\frac{v_F}{2}\int dx\left[\tilde{\Pi}^2+(\partial_x\tilde{\phi})^2\right]\nonumber\\
&+\frac{\lambda_2a}{\pi\xi}\sum_r\cos[2\sqrt{\pi}\tilde{\phi}(r)+4k_Fr],
\end{align}
which can be refermionized by defining the new spinless fermion field $\tilde{\Psi}(x)=e^{i\tilde{k}_Fx}\tilde{\psi}_R(x)+e^{-i\tilde{k}_Fx}\tilde{\psi}_L(x)$ with the slow fields
\begin{align}
\tilde{\psi}_{R,L}(x)=\frac{1}{\sqrt{2\pi\xi}}e^{i\sqrt{\pi}[\pm\tilde{\phi}(x)-\tilde{\vartheta}(x)]},\nonumber
\end{align}
with $\tilde{\Pi}=\partial_x\tilde{\vartheta}$ and $\tilde{k}_F=2k_F$, in terms of which the Hamiltonian reads
\begin{align}
\tilde{H}_c=&-iv_F\int dx\left(\tilde{\psi}^\dag_R\partial_x\tilde{\psi}_R-\tilde{\psi}^\dag_L\partial_x\tilde{\psi}_L\right)\nonumber\\
&+\int dx\left[V^*(x)\tilde{\psi}^\dag_R\tilde{\psi}_L+V(x)\tilde{\psi}^\dag_L\tilde{\psi}_R\right],\nonumber
\end{align}
where $V(x)=e^{2i\tilde{k}_Fx}V_0(x)$ is a single-particle potential with
\begin{align}
V_0(x)=\lambda_2a\sum_{n=-\infty}^\infty\delta(x-na),\nonumber
\end{align}
i.e., a periodic Kronig-Penney potential, $V_0(x)=V_0(x+a)$. The phase $e^{2i\tilde{k}_Fx}$ can be removed from the potential by a chiral rotation $\tilde{\psi}(x)\rightarrow e^{-i\tilde{k}_Fx\gamma^5}\tilde{\psi}(x)$ with $\gamma^5=\sigma_z$ and $\tilde{\psi}=\left(\begin{array}{cc}\tilde{\psi}_R & \tilde{\psi}_L\end{array}\right)^T$. Passing to a first-quantized description, the time-independent Schr\"{o}dinger equation for the single-particle wave function $\tilde{\psi}(x)$ is
\begin{align}
\left[-iv_F\sigma_z\partial_x+V_0(x)\sigma_x\right]\tilde{\psi}=E\tilde{\psi}.\nonumber
\end{align}
This Dirac-Kronig-Penney problem has been studied before\cite{mckellar1987} in the context of one-dimensional quark models of the nucleus, and the single-particle spectrum is given by
\begin{align}\label{LutherEmerySpectrum}
&E_{n,\pm}(k)=\pm\frac{v_F}{a}\cos^{-1}\left[\sech(\pi\rho\lambda_2)\cos ka\right]+2nv_F\frac{\pi}{a},\nonumber\\
&n=0,\pm 1,\pm 2,\ldots,
\end{align}
where $n$ is a band index and the principal branch $0<\cos^{-1}x\leq\pi$ is taken. The crystal momentum $k$ lies in the first Brillouin zone $-\frac{\pi}{a}<k\leq\frac{\pi}{a}$. In the absence of 2-particle backscattering $\lambda_2=0$ we recover the massless Dirac spectrum $E_{0,\pm}(k)=\pm v_F|k|$, while for $\lambda_2\neq 0$ a gap of magnitude $2\Delta$ opens at $k=0$ with $\Delta=v_Fa^{-1}\cos^{-1}[\sech(\pi\rho\lambda_2)]$. In the weak coupling limit $\pi\rho\lambda_2\ll 1$ we have $\Delta\simeq|\lambda_2|$. The low-energy spectrum near the center of the zone ($k\ll \pi/a$) has the form
\begin{align}
E_{0,\pm}(k)=\pm\left(\Delta+\frac{k^2}{2m^*}+\c{O}(k^4)\right),\nonumber
\end{align}
where we define a Newtonian effective mass $m^*=(v_Fa)^{-1}\sinh(\pi\rho|\lambda_2|)$. In the weak coupling limit we have $\Delta\simeq m^*v_F^2$, i.e., a massive Dirac-like spectrum.

The many-body system is gapped only if the Fermi level lies in a gap, which corresponds to integer fillings of the spinless Luther-Emery fermions with respect to the impurity lattice. Because the wave number $\tilde{k}_F$ of the Luther-Emery fermions $\tilde{\psi}$ is twice that of the constituent fermions $\psi$, the many-body system will be gapped only if the constituent fermions are at half-filling with respect to the impurity lattice, i.e., $k_F=\pi/2a$. Therefore, at the Luther-Emery point $K=1/4$ and for half-filling of the conduction electrons with respect to the impurity lattice, the system acquires a gap $\Delta$ for any nonzero $\lambda_2$, where $\Delta\simeq|\lambda_2|$ for small $\rho\lambda_2$. Away from half-filling $k_F\neq\pi/2a$, the system is gapless and is described by a Fermi surface of free Luther-Emery fermions with Fermi wave number $\tilde{k}_F=2k_F$. Since $\tilde{\phi}=2\phi$ and $\tilde{\Pi}=\half\Pi$, we conjecture that these carry charge $2e$ and $S_z$ spin $\hbar/4$.

The Hamiltonian Eq.~(\ref{HcLatticeDecoupling}) can also be studied in the free boson limit $K\ll 1$ which corresponds to very strong electron-electron interactions in the helical liquid. We perform a more general rescaling of the boson fields $\tilde{\Pi}=\sqrt{K}\Pi$ and $\tilde{\phi}=\phi/\sqrt{K}$, of which the Luther-Emery point $K=1/4$ was a special case. At half-filling $k_F=\pi/2a$ and for $K\ll 1$, we can expand $\cos 4\sqrt{\pi K}\tilde{\phi}\simeq-8\pi K\tilde{\phi}^2+\textrm{const.}$ which is appropriate for $\lambda_2<0$ for which $4\sqrt{\pi K}\tilde{\phi}=0(\mod 2\pi)$ in the ground state. Therefore, in the continuum limit the boson $\tilde{\phi}$ develops a mass term $\sim\half M^2\tilde{\phi}^2$ with $M\propto|\lambda_2|^{1/2}$.

It is not difficult to ascertain the physics of the decoupling limit beyond those limiting cases. In the long-wavelength limit $k\ll\pi/a$, we can take the continuum limit $a\rightarrow 0$. Then the discrete sum over impurity sites $r$ in Eq.~(\ref{HcLatticeDecoupling}) can be replaced by an integral over $x$, and we obtain
\begin{align}
\tilde{H}_c=&\frac{v_F}{2}\int dx\left[\tilde{\Pi}^2+(\partial_x\tilde{\phi})^2\right]\nonumber\\
&+\frac{\lambda_2a}{\pi\xi}\int dx\cos(4\sqrt{\pi K}\tilde{\phi}+4k_Fan),\nonumber
\end{align}
where $n\equiv x/a$ is integer (not to be confused with the Luther-Emery band index in Eq.~(\ref{LutherEmerySpectrum})). If $k_F\neq\pi/2a$, the integral over the cosine term averages to zero and we have a free massless boson Hamiltonian with a gapless spectrum. The physics is the same as that of Sec.~\ref{sec:LatticeDecouplingLimit}. At half-filling $k_F=\pi/2a$, we have $4k_Fan=2\pi n$ and the cosine term survives the averaging,
\begin{align}
\tilde{H}_c=\frac{v_F}{2}\int dx\left[\tilde{\Pi}^2+(\partial_x\tilde{\phi})^2\right]
+\frac{\lambda_2a}{\pi\xi}\int dx\cos 4\sqrt{\pi K}\tilde{\phi},\nonumber
\end{align}
i.e., the usual sine-Gordon model. A term of the form $\cos\beta\tilde{\phi}$ is relevant in the infrared\cite{Fradkin} for $\beta^2<8\pi$. Here we have $\beta^2=16\pi K$. On the one hand, if $K>1/2$ the cosine term is irrelevant. The spin $U(1)_s$ symmetry is dynamically restored at low energies, the conduction electrons remain gapless, the field $\phi$ fluctuates wildy and there is no long-range order of the impurity spins. Once again, the physics is the same as that of Sec.~\ref{sec:LatticeDecouplingLimit}. On the other hand, if $K<1/2$ the cosine term is relevant. The ground state spontaneously breaks the $\mathbb{Z}_2$ spin symmetry with long-range order of the field $\phi$. The Luther-Emery and free boson points are particular points in that phase. The conduction electrons open up a gap\cite{Wu2006} $\Delta\propto|\lambda_2|^{1/(2-4K)}$ for $\rho\lambda_2\ll 1$, which agrees with the expressions $\Delta\propto|\lambda_2|$ and $\Delta\propto|\lambda_2|^{1/2}$ in the Luther-Emery $K=1/4$ and free boson $K\rightarrow 0$ limits, respectively. As in Sec.~\ref{sec:SingleImp2PB}, the local CDW phase $2\sqrt{\pi}\phi(r)$ is pinned in the ground state at $(n+\half)\pi$ for $\lambda_2>0$ and $n\pi$ for $\lambda_2<0$ with $n\in\mathbb{Z}$. To the difference of the single-impurity case however, here there is long-range spatial order of the CDW phase. Using Eq.~(\ref{LatticeCDWUnitVector}) and Eq.~(\ref{LocalCDWOrderLattice}) one can see that the impurity spins develop long-range Ising antiferromagnetic order,
\begin{align}
\langle S^y_r\rangle&=\pm\half(-1)^{r/a}\sgn J_\perp,\hspace{5mm}\lambda_2>0,\nonumber\\
\langle S^x_r\rangle&=\pm\half(-1)^{r/a}\sgn J_\perp,\hspace{5mm}\lambda_2<0,\nonumber
\end{align}
i.e., the order is either in the $y$ or $x$ direction depending on the sign of $\lambda_2$. The $\pm$ sign corresponds to the two degenerate antiferromagnetic ground states and is picked by spontaneous symmetry breaking.

As in Sec.~\ref{sec:SingleImp2PB}, one can ask whether the scaling dimension of the 2-particle backscattering term at half-filling is affected by the impurity spin sector away from the decoupling limit. It is still true that the 2-particle backscattering operator commutes with the unitary transformation $U=\exp(i\lambda\sum_r\phi(r)S^z_r)$ with $\lambda=-J_za/\sqrt{\pi}v_FK$ for all $J_z$ and hence that the correlator $\langle\cos 4\sqrt{\pi}\phi(x,\tau)\cos 4\sqrt{\pi}\phi(0,0)\rangle$ is independent of $J_z$. Therefore the scaling dimension of the 2-particle backscattering operator is still $4K$. However, in the Kondo lattice problem, the bulk Luttinger parameter $K$ does renormalize under the $(1+1)$-dimensional RG flow. In particular, $K$ is renormalized by the impurity spin sector even in the absence of 2-particle backscattering, as the Kosterlitz-Thouless equations (\ref{KTeq1}) and (\ref{KTeq2}) show. One could therefore expect that the phase boundary at $K=1/2$ is changed by the presence of the Kondo lattice.

To check whether this is the case or not, we repeat the analysis of Sec.~\ref{sec:2DCG} in the presence of the 2-particle backscattering term $S_\mathrm{2PB}=\frac{\lambda_2a}{\pi\xi}\int_0^\infty d\tau\sum_r\cos[4\sqrt{\pi}\phi(r,\tau)+4k_Fr]$. The technical details of the mapping to a Coulomb gas are similar and will not be reproduced here. The main differences with the $\lambda_2=0$ problem are as follows. In addition to the Ising variables $m_\ell=\pm 1$ representing spin flips, we need to introduce another set of Ising variables $e_j=\pm 1$ representing 2-particle backscattering events. Because two spin flips also backscatter two conduction electrons, the dynamics of the $m_\ell$ and $e_j$ particles are coupled. After neglecting the noninteracting factors $Z_\mathrm{TL}$ and $Z_S$ (see discussion following Eq.~(\ref{ZCG})), the zero temperature partition function of the Kondo lattice in the presence of 2-particle backscattering is exactly mapped to that of two coupled 2D classical Coulomb gases,\cite{nelson1980,cardy1982}
\begin{widetext}
\begin{align}\label{ZCG2PB}
Z=\sum_{N_m=0}^\infty\sum_{N_e=0}^\infty\sum_{\{m\}}'\sum_{\{e\}}'\frac{Y_m^{N_m}}{N_m!}\frac{Y_e^{N_e}}{N_e!}
\int_{|\Delta\b{x}^e|>\xi}\mathcal{D}\b{x}^e\int_{|\Delta\b{x}^m|>\xi}\mathcal{D}\b{x}^m\,e^{-S_\mathrm{CG}},
\end{align}
%\begin{align}\label{ZCG2PB}
%Z=&\sum_{N_m=0}^\infty\sum_{N_e=0}^\infty\sum_{\{m\}}'\sum_{\{e\}}'\frac{Y_m^{N_m}}{N_m!}\frac{Y_e^{N_e}}{N_e!}\nonumber\\
%&\times\int_{|\Delta\b{x}^e|>\xi}\mathcal{D}\b{x}^e\int_{|\Delta\b{x}^m|>\xi}\mathcal{D}\b{x}^m\,e^{-S_\mathrm{CG}},
%\end{align}
where $N_m$ and $Y_m=\rho|J_\perp|/2\sqrt{2}$ are the total number of spin flips and their fugacity (i.e., the same as $N$ and $Y$ in Eq.~(\ref{ZCG})), $N_e$ and $Y_e=\rho|\lambda_2|/2$ are the total number of 2-particle backscattering events and their fugacity, the primed sum over spin flip configurations $\{m\}$ is the same as that in Eq.~(\ref{ZCG}), the primed sum over 2-particle backscattering event configurations $\{e\}$ is subject to the global neutrality constraint $\sum_j e_j=0$, and the integration measures $\int\mathcal{D}\b{x}^m$ and $\int\mathcal{D}\b{x}^e$ are as in Eq.~(\ref{CoulombGasIntegrationMeasure}), subject to the hard-core constraints $|\b{x}^m_\ell-\b{x}^m_{\ell'}|>\xi$, $\ell\neq\ell'$ and $|\b{x}^e_j-\b{x}^e_{j'}|>\xi$, $j\neq j'$, for particles of same type. The action in Eq.~(\ref{ZCG2PB}) is
%\begin{widetext}
\begin{align}\label{SCG2PB}
S_\mathrm{CG}=-\frac{1}{2}\left(g_{mm}\sum_{\ell\neq\ell'}m_\ell m_{\ell'}\ln\frac{|\b{x}^m_\ell-\b{x}^m_{\ell'}|}{\xi}
+g_{ee}\sum_{j\neq j'}e_j e_{j'}\ln\frac{|\b{x}^e_j-\b{x}^e_{j'}|}{\xi}
+g_{em}\sum_{j\ell}e_jm_\ell\ln\frac{|\b{x}^e_j-\b{x}^m_\ell|}{\xi}\right)
+4ik_F\sum_j r_j^e e_j,
\end{align}
\end{widetext}
where $g_{mm}=2K\chi^2$ (i.e., the same as $g$ in Eq.~(\ref{SCGno2PB})), $g_{ee}=8K$, and $g_{em}=8K\chi$. The last term in Eq.~(\ref{SCG2PB}) is pure imaginary and can be thought of as a Berry phase effect, with $r^e_j$ the spatial coordinate of the $j$th 2-particle backscattering event (recall that $\b{x}=(r,v_F\tau)$). Unlike for the spin flips, there is no local neutrality condition which would allow us to set this term to zero. However, due to the presence of the impurity lattice $r^e_j$ is an integer multiple of $a$. Furthermore, $e_j=\pm 1$ is an integer. Therefore, at half-filling $2k_F=\pi/a$ we find that $4k_F\sum_j r_j^e e_j$ is an integer multiple of $2\pi$, hence $\exp(-4ik_F\sum_j r_j^e e_j)=1$ and the Berry phase term does not contribute to the partition function. Away from half-filling, this complex factor is oscillatory and strongly suppresses configurations of $e$ particles. Only the trivial configuration with $N_e=0$, i.e., with no $e$ particles whatsoever, survives the partition sum. In other words, the fugacity $Y_e\propto|\lambda_2|$ becomes an irrelevant variable under the RG. We thus confirm the result expected from the analysis in the decoupling limit that the 2-particle backscattering term is irrelevant away from half-filling, regardless of the value of $J_z$.

In the following we focus on the half-filled case $2k_F=\pi/a$ where both $m$ and $e$ particles need to be retained, and the action is given by Eq.~(\ref{SCG2PB}) without the Berry phase term. The scaling dimensions of the fugacity variables $Y_e$ and $Y_m$ can be obtained by rescaling the cutoff $\xi\rightarrow\xi+d\xi$ in the integration measures $\int\mathcal{D}\b{x}$ and in the logarithmic interaction potentials $\ln(|\b{x}-\b{x}'|/\xi)$.\cite{nienhuis} In general, to find all the relevant variables one also needs to consider the fugacities of new particle types which are not present in the original problem but are created along the RG flow by particle ``fusion''.\cite{nienhuis} Those particles are generally of the type $(p,q)$ which means that they are composite objects of $p$ particles of type $m$ and $q$ particles of type $e$. The scaling dimensions of all possible such particles are indicated in Table~\ref{table:scalingdim}.
\begin{table}
\begin{tabular}{@{\extracolsep{1cm}}cc}
  \hline\hline
  Particle type $a$ & Scaling dimension $\Delta_a$ \\ \hline
  $(1,0)\equiv m$ & $K\chi^2$  \\
  $(2,0)$ & $4K\chi^2$ \\
  \vdots & \vdots \\
  $(p,0)$ & $K\chi^2p^2$ \\ \hline
  $(0,1)\equiv e$ & $4K$  \\
  $(0,2)$ & $16K$ \\
  \vdots & \vdots \\
  $(0,q)$ & $4Kq^2$ \\   \hline
  $(1,1)$ & $K\chi^2+4K$ \\
  \vdots & \vdots \\
  $(p,q)$ & $K\chi^2p^2+4Kq^2$ \\ \hline\hline
\end{tabular}
\caption{New particles generated under renormalization and their scaling dimensions.}
\label{table:scalingdim}
\end{table}
A particle of type $a$ is relevant if its scaling dimension $\Delta_a<2$. For $K>1/2$, all particles of the type $(p,q>0)$ are irrelevant, i.e., only pure spin flips are relevant.

What about the renormalization of $K$ due to spin flips? This effect which corresponds here to the renormalization of $g_{ee}$ occurs at one-loop level. To illustrate the physics, we compute the one-loop RG equations taking into account the $e$ and $m$ particles, as well as the next most relevant composite spin flip operator $(2,0)\equiv\tilde{m}$ which corresponds to double spin flips. We find
\begin{align}
\frac{dY_e}{d\ell}&=\left(2-\half g_{ee}\right)Y_e,\nonumber\\
\frac{dY_m}{d\ell}&=\left(2-\half g_{mm}\right)Y_m+2\pi Y_mY_{\tilde{m}},\nonumber\\
\frac{dY_{\tilde{m}}}{d\ell}&=(2-2g_{mm})Y_{\tilde{m}}+\pi Y_m^2,\nonumber\\
\frac{dg_{ee}}{d\ell}&=-2\pi^2\left[Y_e^2g_{ee}^2+\left(Y_m^2+4Y_{\tilde{m}}^2\right)g_{em}^2\right],\nonumber\\
\frac{dg_{mm}}{d\ell}&=-2\pi^2\left[\left(Y_m^2+4Y_{\tilde{m}}^2\right)g_{mm}^2+Y_e^2g_{em}^2\right],\nonumber\\
\frac{dg_{em}}{d\ell}&=-2\pi^2\left[Y_e^2g_{ee}+\left(Y_m^2+4Y_{\tilde{m}}^2\right)g_{mm}\right]g_{em}.\nonumber
\end{align}
The beta function of $g_{ee}$ is negative, which means that $Y_e$ tends to be more relevant as we flow into the infrared than if there was no renormalization of $g_{ee}$. However, the beta function of $g_{ee}$ is second order in all the fugacities, which for small $J_\perp$ and $\lambda_2$ are small. The consideration of higher order spin flips $(p>2,0)$ will give additional contributions to the beta function of $g_{ee}$ but they will all be quadratic in the corresponding fugacities, and therefore small for small $J_\perp$ and $\lambda_2$. We therefore conclude that for small $J_\perp$ and $\lambda_2$, the phase boundary between the gapped Ising antiferromagnet and the gapless disordered state does indeed occur at $K=1/2$ regardless of the value of $J_z$. We expect that the phase boundary will be affected by the one-loop RG flows for finite $J_\perp$ and $\lambda_2$, but this is a regime where the perturbative approach described here eventually fails. For the same reasons, we also expect that the lattice version $\sum_r[(\partial_x^2\vartheta(r))e^{i2\sqrt{\pi}\phi(r)}+\mathrm{h.c.}]$ of the inelastic backscattering operator\cite{schmidt2011,lezmy2012} mentioned in Sec.~\ref{sec:SingleImp2PB} remains irrelevant for all $0<K<1$.

\subsection{Phase diagram of the Kondo lattice problem}

\begin{figure}[t]
\begin{center}
\includegraphics[width=3.5in,angle=0]{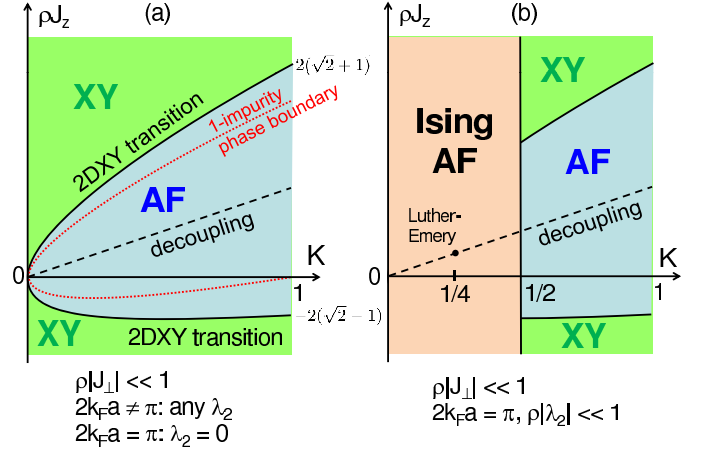}
\end{center}
\caption{Zero temperature phase diagram of the 1D Kondo lattice in the helical edge liquid of the quantum spin Hall state, for small $J_\perp$. (a) In the absence of 2-particle backscattering for any filling, or in the presence of 2-particle backscattering but away from half-filling ($2k_Fa\neq\pi$), there are two gapless phases: a Kondo screened strong coupling phase (AF) and an unscreened XY phase, separated by a quantum phase transition in the 2DXY universality class. The decoupling limit (dashed line) lies in the AF phase. The single-impurity phase boundary of Fig.~\ref{fig:fig2}(b) is drawn for comparison (dotted line). (b) In the presence of 2-particle backscattering and at half-filling ($2k_Fa=\pi$), there is an additional gapped phase with long-range Ising antiferromagnetic order (Ising AF) for $K<1/2$. In contrast to the ordinary half-filled 1D Kondo lattice (Ref.~\onlinecite{tsunetsugu1997}), for noninteracting conduction electrons ($K=1$) the system remains gapless.}
\label{fig:PhaseDiagramKL}
\end{figure}

We now discuss the phase diagram of the Kondo lattice problem based on the results obtained thus far (Fig.~\ref{fig:PhaseDiagramKL}). There are four cases to distinguish: at half-filling ($2k_Fa=\pi$) or away from half-filling ($2k_Fa\neq\pi$), with 2-particle backscattering ($\lambda_2\neq 0$) or without ($\lambda_2=0$). In three out of these four cases (away from half-filling without 2-particle backscattering, away from half-filling with 2-particle backscattering, and at half-filling without 2-particle backscattering) the phase diagram is given in Fig.~\ref{fig:PhaseDiagramKL}(a). There is a Kondo screened strong coupling phase (AF) sandwiched between two XY phases. In the AF phase, the spin flip fugacity $Y_m$ and hence $|J_\perp|$ flow to infinity. Based on the exact solution in the decoupling limit (Sec.~\ref{sec:LatticeDecouplingLimit}), we expect that impurity spin correlations are gapless in the $xy$ plane but gapped in the out-of-plane $z$ direction with a gap of order $|J_\perp|$. In the XY phase, $Y_m$ flows to zero and the system is described by an effective Hamiltonian similar to Eq.~(\ref{HKL}) but with $J_\perp=0$ and renormalized values of $K$ and $J_z$. As in the single-impurity case (Sec.~\ref{sec:SingleImpAYH}), the impurity spin correlations can be computed in this phase by applying the unitary transformation $U$ on the lattice to remove the $\sum_rS^z\Pi(r)$ term from the effective Hamiltonian; they are gapless in the $xy$ plane and vanish in the $z$ direction. Although the phase diagram of the Kondo lattice is qualitatively similar to that of the single-impurity problem in Fig.~\ref{fig:fig2}(b), one interesting difference is that for noninteracting conduction electrons ($K=1$), a ferromagnetic $J_z<0$ can give rise to a strong coupling AF phase at low energies, which did not occur for a single impurity.

At half-filling and in the presence of 2-particle backscattering, there is an additional gapped phase with long-range Ising antiferromagnetic order (Ising AF) for $K<1/2$ [Fig.~\ref{fig:PhaseDiagramKL}(b)]. The Luther-Emery point at $K=1/4$ and $\rho J_z=2K=1/2$ is one point in that phase. For $K>1/2$, 2-particle backscattering is irrelevant and that portion of the phase diagram is the same as Fig.~\ref{fig:PhaseDiagramKL}(a). In particular, for noninteracting conduction electrons ($K=1$) the system remains gapless, being either in the AF or the XY phase. This is in marked contrast to the ordinary 1D Kondo lattice with non-helical spinful conduction electrons, in which case the ground state of the system is a gapped spin liquid.\cite{tsunetsugu1992,tsvelik1994,tsunetsugu1997} The Kondo lattice in a helical liquid can open a gap, but only for $K<1/2$ and at the cost of developing Ising antiferromagnetic long-range order, which is also qualitatively different from the usual spinful case. The Mermin-Wagner theorem, which forbids the existence of long-range magnetic order in quantum $(1+1)$-dimensional spin systems with a continuous spin rotation symmetry, has recently been extended to the case of lattice spins coupled to itinerant charge carriers such as the 1D Kondo lattice.\cite{loss2011} The disordered ground state of the half-filled $SU(2)$-invariant 1D Kondo lattice is a good example of this general result. However, in the presence of spin-orbit interactions magnetic order is not excluded.\cite{loss2011} In our case, the 2-particle backscattering operator breaks the continuous $U(1)$ spin rotation symmetry to the discrete $\mathbb{Z}_2$ symmetry, and microscopically arises from electron-electron interactions in the presence of spin-orbit coupling.\cite{Wu2006} This allows the system to escape the extended Mermin-Wagner theorem and develop long-range Ising AF order.

\section{Conclusion}

We have generalized previous studies of quantum impurities on the 1D edge of a 2D time-reversal invariant topological insulator (QSH insulator) in two directions.

First, we derived the zero temperature phase diagram of the single-impurity problem in the $(K,J_z)$ plane for all values $0<K<1$ of the Luttinger parameter $K$ of the helical edge liquid, corresponding to repulsive electron-electron interactions, and all values of the Kondo coupling $J_z$ in the $z$ direction. Previous treatments were restricted to the weak coupling regime $1-K\ll 1$ and $\rho|J_z|\ll 1$. We found that a large portion of the phase diagram for strong antiferromagnetic $J_z>0$ was occupied by a local moment phase which usually occurs only for ferromagnetic $J_z<0$. This result had been missed by previous works. Our new results were derived in part by making use of an exact solution in the so-called decoupling limit $\rho J_z=2K$ which corresponds, at least in the noninteracting limit $K=1$, to the unitarity limit $\delta=\pm\frac{\pi}{2}$ for the scattering of the edge electrons on the $z$ component of the impurity spin. The solution in the decoupling limit was exact for all $J_\perp$. This analysis was supplemented by a calculation \`a la Anderson-Yuval-Hamann which allowed us to derive improved renormalization group equations for the Kondo couplings, and to explore the phase diagram away from the decoupling limit.

Second, we generalized the single-impurity problem to a Kondo lattice problem where a regular 1D array of quantum impurities interacted with the edge electrons. The solution in the decoupling limit was extended to the lattice. We found that the topology of the zero temperature phase diagram was similar to that of the single-impurity problem. However, an interesting difference was that in the noninteracting case ($K=1$), ferromagnetic Kondo couplings could give rise to a Kondo screened phase in the lattice case but not in the single-impurity case. More importantly, the physics of the Kondo lattice problem was found to depend crucially on the filling of conduction electrons with respect to the impurity lattice. Away from half-filling, we found two gapless phases separated by a quantum phase transition in the 2DXY universality class. At half-filling, we found an additional gapped phase for $K<1/2$ with long-range Ising antiferromagnetic order. This was contrasted with the disordered ground state of the half-filled Kondo lattice in an ordinary spinful 1D electron gas.

\acknowledgments
We thank E. Berg, J. L. Cardy, B. K. Clark, B. I. Halperin, D. A. Huse, S. A. Kivelson, D. R. Nelson, S. Ostlund, M. Schir\'{o}, H. L. Verlinde, and M. Yamazaki for useful discussions. This research was supported by the Simons Foundation and by the National Science Foundation under Grant No.~NSF PHY05-51164. We acknowledge the hospitality of the Kavli Institute for Theoretical Physics (KITP) were part of this work was completed.

\appendix
\section{Decoupling limit and unitarity limit}

In this Appendix we show that for noninteracting edge electrons ($K=1$), the decoupling limit $\rho J_z=2K=2$ corresponds to the unitarity limit $\delta=\pm\frac{\pi}{2}$ for scattering on the $z$ component of the impurity spin, where $\delta$ is the scattering phase shift. For noninteracting electrons, we can work in the fermion representation where the single-impurity Hamiltonian Eq.~(\ref{HSingleImp}) in the absence of the $J_\perp$ term is
\begin{align}
H=&-iv_F\int dx\left(\psi_{R\uparrow}^\dag\partial_x\psi_{R\uparrow}-\psi_{L\downarrow}^\dag\partial_x\psi_{L\downarrow}\right)\nonumber\\
&+J_zaS^z\left(\psi_{R\uparrow}^\dag\psi_{R\uparrow}-\psi_{L\downarrow}^\dag\psi_{L\downarrow}\right)_{x=0},\nonumber
\end{align}
where $\psi_{R\uparrow}^\dag$ and $\psi_{L\downarrow}^\dag$ are the fermionic creation operators for the spin-up right-moving edge electrons and spin-down left-moving edge electrons, respectively. Since $[H,S^z]=0$, we can work in the $S^z$ basis and treat $S^z=\pm\half$ as a $c$-number. Furthermore, $H$ is also diagonal in the electron spin $s^z$ basis, $H=\sum_{\sigma=\pm 1}\int dx\,\psi^\dag_\sigma\hat{H}_\sigma\psi_\sigma$ where the first-quantized Hamiltonian $\hat{H}_\sigma$ is
\begin{align}
\hat{H}_\sigma=-i\sigma v_F\partial_x+\sigma J_zaS^z\delta(x).\nonumber
\end{align}
The time-independent Schr\"odinger equation $\hat{H}_\sigma\varphi=E\varphi$ is solved by the ansatz $\varphi_{k\sigma}(x)\sim e^{i\sigma kx+if(x)}$ with $f$ such that $\partial_xf=0$ for $x\neq 0$. For $x\neq 0$, the Schr\"odinger equation reads $-i\sigma v_F\partial_x\varphi_{k\sigma}=E\varphi_{k\sigma}$ which gives the expected linear dispersion $E=v_Fk$. Substituting this result in the Schr\"odinger equation at $x=0$, we obtain $v_F\partial_xf+J_zaS^z\delta(x)=0$ which is solved by $f(x)=-\pi\rho J_zS^z\theta(x)$ where $\rho=a/\pi v_F$ is the density of states of the helical liquid and $\theta(x)$ is the Heaviside step function. As a result, the edge state wave function is
\begin{align}
\varphi_{k\sigma}(x)\sim e^{i\sigma kx-i\pi\rho J_zS^z\theta(x)}.\nonumber
\end{align}
The scattering $S$-matrix is defined by
\begin{align}
S(k)=e^{2i\delta_k}=\frac{\varphi_{k\sigma}(x=0^+)}{\varphi_{k\sigma}(x=0^-)}=e^{i\pi\rho J_zS^z},\nonumber
\end{align}
hence the scattering phase shift $\delta_k=\delta$ is
\begin{align}
\delta=-\half\pi\rho J_zS^z=\pm{\textstyle\frac{1}{4}}\pi\rho J_z,\nonumber
\end{align}
since $S^z=\pm\half$. In the decoupling limit, we have $\rho J_z=2$ which corresponds to the unitarity limit $\delta=\pm\frac{\pi}{2}$.

\bibliography{kondolattice}
\end{document}